\documentclass[11pt]{article}
\usepackage{amssymb,amsmath,amsfonts}
\usepackage{graphicx}
\usepackage{graphics}
\usepackage{eepic,epsfig}

1.60cm \makeatletter \@addtoreset{equation}{section}

\makeatother

\begin{document}

\title{Wightman function and vacuum fluctuations in higher dimensional brane
models}
\author{Aram A. Saharian\thanks{%
Email: saharyan@server.physdep.r.am} \\
\textit{Department of Physics, Yerevan State University, 1 Alex Manoogian
Str.} \\
\textit{\ 375049 Yerevan, Armenia,}\\
\textit{and}\\
\textit{The Abdus Salam International Centre for Theoretical Physics} \\
\textit{\ 34014 Trieste, Italy }}
\date{\today}
\maketitle

\begin{abstract}
Wightman function and vacuum expectation value of the field square are
evaluated for a massive scalar field with general curvature coupling
parameter subject to Robin boundary conditions on two codimension one
parallel branes located on $(D+1)$-dimensional background spacetime $%
AdS_{D_1+1}\times \Sigma $ with a warped internal space $\Sigma $. The
general case of different Robin coefficients on separate branes is
considered. The application of the generalized Abel-Plana formula for the
series over zeros of combinations of cylinder functions allows us to extract
manifestly the part due to the bulk without boundaries. Unlike to the purely
AdS bulk, the vacuum expectation value of the field square induced by a
single brane, in addition to the distance from the brane, depends also on
the position of the brane in the bulk. The brane induced part in this
expectation value vanishes when the brane position tends to the AdS horizon
or AdS boundary. The asymptotic behavior of the vacuum densities near the
branes and at large distances is investigated. The contribution of
Kaluza-Klein modes along $\Sigma $ is discussed in various limiting cases.
In the limit when the curvature radius for the AdS spacetime tends to
infinity, we derive the results for two parallel Robin plates on background
spacetime $R^{(D_1,1)}\times \Sigma $. For strong gravitational fields
corresponding to large values of the AdS energy scale, the both single brane
and interference parts of the expectation values integrated over the
internal space are exponentially suppressed. As an example the case $\Sigma
=S^1$ is considered, corresponding to the $AdS_{D+1}$ bulk with one
compactified dimension. An application to the higher dimensional
generalization of the Randall-Sundrum brane model with arbitrary mass terms
on the branes is discussed.
\end{abstract}

\bigskip


\bigskip

\section{Introduction}

\label{sec:introd}

Anti-de Sitter (AdS) spacetime is remarkable from different points of view.
The early interest to this spacetime was motivated by the questions of
principal nature related to the quantization of fields propagating on curved
backgrounds. The investigation of the dynamics of fields on AdS is of
interest not only because AdS space is a space of high symmetry, and hence
exact solutions for the free field theory can be written down, but also
because it is a space of constant nonzero curvature, and thus field theory
in AdS background is not a trivial rewriting of Minkowski spacetime field
theory. The presence of the both regular and irregular modes and the
possibility of interesting causal structure lead to many new phenomena. The
importance of this theoretical work increased when it has been discovered
that AdS spacetime generically arises as ground state in extended
supergravity and in string theories. Further interest in this subject was
generated by the appearance of two models where AdS geometry plays a special
role. The first model, the so called AdS/CFT correspondence (for a review
see \cite{Ahar00}), represents a realization of the holographic principle
and relates string theories or supergravity in the AdS bulk with a conformal
field theory living on its boundary. It has many interesting consequences
and provides a powerful tool to investigate gauge field theories. The second
model is a realization of a braneworld scenario with large extra dimensions
and provides a solution to the hierarchy problem between the gravitational
and electroweak mass scales (for reviews in braneworld gravity and cosmology
see Ref. \cite{Ruba01}). In this model the main idea to resolve the large
hierarchy is that the small coupling of four dimensional gravity is
generated by the large physical volume of extra dimensions. Braneworlds
naturally appear in string/M-theory context and provide a novel setting for
discussing phenomenological and cosmological issues related to extra
dimensions. The model introduced by Randall and Sundrum \cite{Rand99a} is
particularly attractive. The corresponding background solution consists of
two parallel flat branes, one with positive tension and another with
negative tension embedded in a five dimensional AdS bulk. The fifth
coordinate is compactified on $S^1/Z_2$, and the branes are on the two fixed
points. In the original version of the model it is assumed that all matter
fields are confined on the branes and only the gravity propagates freely in
the five dimensional bulk. More recently, alternatives to confining
particles on the brane have been investigated and scenarios with additional
bulk fields have been considered. Apart from the hierarchy problem it has
been also tried to solve the cosmological constant problem within the
braneworld scenario. The braneworld theories may give some alternative
discussion of the cosmological constant. The basic new ingredient is that
the vacuum energy generated by quantum fluctuations of fields living on the
brane may not curve the brane itself but instead the space transverse to it.

Randall-Sundrum scenario is just the simplest possibility within a more
general class of higher dimensional warped geometries. Such a generalization
is of importance from the viewpoint of a underlying fundamental theory in
higher dimensions such as ten dimensional superstring theory. From a
phenomenological point of view, higher dimensional theories with curved
internal manifolds offer a richer geometrical and topological structure. The
consideration of more general spacetimes may provide interesting extensions
of the Randall-Sundrum mechanism for the geometric origin of the hierarchy.
Spacetimes with more than one extra dimension can allow for solutions with
more appealing features, particularly in spacetimes where the curvature of
the internal space is non-zero. More extra dimensions also relax the
fine-tunings of the fundamental parameters. These models can provide a
framework in the context of which the stabilization of the radion field
naturally takes place. Several variants of the Randall--Sundrum scenario
involving cosmic strings and other global defects of various codimensions
has been investigated in higher dimensions (see, for instance, \cite{Greg00}-%
\cite{Davo03} and references therein). In particular, much work has been
devoted to warped geometries in six dimensions (see references in \cite%
{Kofi05}).

In braneworld models the investigation of quantum effects is of considerable
phenomenological interest, both in particle physics and in cosmology. The
braneworld corresponds to a manifold with boundaries and all fields which
propagate in the bulk will give Casimir-type contributions to the vacuum
energy (for reviews of the Casimir effect see Ref. \cite{Most97}), and as a
result to the vacuum forces acting on the branes. Casimir forces provide a
natural alternative to the Goldberger-Wise mechanism for stabilizing the
radion field in the Randall-Sundrum model, as required for a complete
solution of the hierarchy problem. In addition, the Casimir energy gives a
contribution to both the brane and bulk cosmological constants and, hence,
has to be taken into account in the self-consistent formulation of the
braneworld dynamics. Motivated by these, the role of quantum effects in
braneworld scenarios has received a great deal of attention \cite{Fabi00}-%
\cite{Pujo05}. However, in the most part of these papers the
authors consider the global quantities such as the total Casimir
energy, effective action or conformally invariant fields. The
investigation of local physical characteristics in the Casimir
effect, such as expectation value of the energy-momentum tensor
and the field square, is of considerable interest. Local
quantities contain more information on the vacuum fluctuations
than the global ones. In addition to describing the physical
structure of the quantum field at a given point, the
energy-momentum tensor acts as the source in the Einstein
equations and therefore plays an important role in modelling a
self-consistent dynamics involving the gravitational field.
Quantum fluctuations play also an important role in inflationary
cosmology
as they are related to the power spectrum. As it has been shown in Ref. \cite%
{Pujo05}, the quantum fluctuations of a bulk scalar field coupled to a brane
located scalar field with a bi-quadratic interaction generate an effective
potential for the field on the brane with a true vacuum at the nonzero
values of the field. In particular, these calculations are relevant to the
bulk inflaton model \cite{Koba01}, where the inflation on the brane is
driven by the bulk scalar field. In the case of two parallel branes on AdS
background, the vacuum expectation value of the bulk energy-momentum tensor
for a scalar field with an arbitrary curvature coupling is investigated in
Refs. \cite{Knap03,Saha04a}. In particular, in Ref. \cite{Saha04a} the
application of the generalized Abel-Plana formula \cite{Sahreview,Sahsph} to
the corresponding mode sums allowed us to extract manifestly the parts due
to the AdS spacetime without boundaries and to present the boundary induced
parts in terms of exponentially convergent integrals for the points away the
boundaries. The interaction forces between the branes are investigated as
well. Depending on the coefficients in the boundary conditions, these forces
can be either attractive or repulsive. The local vacuum effects for a bulk
scalar field in brane models with dS branes are discussed in Refs. \cite%
{Pujo04,Nayl05}. On background of manifolds with boundaries, the physical
quantities, in general, will receive both volume and surface contributions.
For scalar fields with general curvature coupling, in Ref. \cite{Rome02} it
has been shown that in the discussion of the relation between the mode sum
energy, evaluated as the sum of the zero-point energies for each normal mode
of frequency, and the volume integral of the renormalized energy density for
the Robin parallel plates geometry it is necessary to include in the energy
a surface term located on the boundary. An expression for the surface
energy-momentum tensor for a scalar field with general curvature coupling
parameter in the general case of bulk and boundary geometries is derived in
Ref. \cite{Saha03}. The corresponding vacuum expectation values are
investigated in Ref. \cite{Saha04} for the model with two flat parallel
branes on AdS bulk and in \cite{Pujo04} for a dS brane in the flat bulk. In
particular, for the first case it has been shown that in the Randall-Sundrum
model the cosmological constant induced on the visible brane by the presence
of the hidden brane is of the right order of magnitude with the value
suggested by the cosmological observations without an additional fine tuning
of the parameters.

The purpose of the present paper is to study the Wightman function and the
vacuum expectation value of the field square for a scalar field with an
arbitrary curvature coupling parameter obeying Robin boundary conditions on
two codimension one parallel branes embedded in the background spacetime $%
AdS_{D_1+1}\times \Sigma $ with a warped internal space $\Sigma $. The
quantum effective potential and the problem of moduli stabilization in the
orbifolded version of this model with zero mass parameters on the branes are
discussed recently in Ref. \cite{Flac03b}. In particular, it has been shown
that one loop-effects induced by bulk scalar fields generate a suitable
effective potential which can stabilize the hierarchy without fine tuning.
The corresponding results are extended to the type of models with unwarped
internal space $\Sigma $ in Ref. \cite{Flac03}. We organize the present
paper as follows. In the next section we evaluate the Wightman function in
the region between the branes. By using the generalized Abel-Plana formula,
we present this function in the form of a sum of the Wightman function for
the bulk without boundaries and boundary induced parts. The vacuum
expectation value of the field square for a general case of the internal
space $\Sigma $ is discussed in section \ref{sec:vevphi2} for the case of a
single brane geometry and in section \ref{sec:phi2twopl} in the region
between two branes. Various limiting cases are discussed when the general
formulae are simplified. A simple example with the internal space $S^1$ is
then considered in section \ref{sec:example}. The last section contains a
summary of the work. The vacuum expectation values of the energy-momentum
tensor and the interaction forces between the branes will be discussed in
the forthcoming paper.

\section{Wightman function}

\label{sec:WF}

Consider a scalar field $\varphi (x)$ with curvature coupling parameter $%
\zeta $ satisfying the equation of motion
\begin{equation}
\left( g^{MN}\nabla _{M}\nabla _{N}+m^{2}+\zeta R\right) \varphi (x)=0,
\label{fieldeq}
\end{equation}%
$M,N=0,1,\ldots ,D$, with $R$ being the scalar curvature for a $(D+1)$%
-dimensional background spacetime, $\nabla _{M}$ is the covariant derivative
operator associated with the metric tensor $g_{MN}$ (we adopt the
conventions of Ref. \cite{Birrell} for the metric signature and the
curvature tensor). For minimally and conformally coupled scalars one has $%
\zeta =0$ and $\zeta =\zeta _{D}\equiv (D-1)/4D$ correspondingly. We will
assume that the background spacetime has a topology $AdS_{D_1+1}\times
\Sigma $, where $\Sigma $ is a $D_2$-dimensional manifold.

First let us consider a general class of spacetimes described by the line
element (see also the discussion in Ref. \cite{Flac03b})
\begin{equation}
ds^{2}=g_{MN}dx^{M}dx^{N}=e^{-2\sigma (y)}\eta _{\mu \nu }dx^{\mu }dx^{\nu
}-e^{-2\rho (y)}\gamma _{ij}dX^{i}dX^{j}-dy^{2},  \label{metric}
\end{equation}%
with $\eta _{\mu \nu }=\mathrm{diag}(1,-1,\ldots ,-1)$ being the metric for
the $D_{1}$-dimensional Minkowski spacetime $R^{(D_1-1,1)}$ and the
coordinates $X^{i}$ cover the manifold $\Sigma $, $ D=D_{1}+D_{2}$. Here and
below $\mu ,\nu =0,1,\ldots ,D_{1}-1$ and $i,j=1,\ldots ,D_{2}$. The scalar
curvature for the metric tensor from (\ref{metric}) is given by the
expression
\begin{equation}  \label{Ricci}
R=-F(y)-e^{2\rho }R_{(\gamma )},
\end{equation}
where we have introduced the notation
\begin{equation}  \label{Fy}
F(y)=-2\tau ^{\prime \prime }(y)+\tau ^{\prime 2}(y)+D_{1}\sigma ^{\prime
2}(y)+D_{2}\rho ^{\prime 2}(y),
\end{equation}
with $\tau (y)=D_{1}\sigma (y)+D_{2}\rho (y)$ and $R_{(\gamma )}$ is the
scalar curvature for the metric tensor $\gamma _{ik}$. The background metric
for many braneworld models falls into the general class corresponding to Eq.
(\ref{metric}). In particular, the metric with $D_2=0$ and $\sigma (y)=k_D|y|
$ is the Randall-Sundrum case. The metric solutions with $D_2=1$, $\sigma
(y)=k_D|y|$, $\rho (y)={\mathrm{const}}$ and $D_2=1$, $\sigma (y)=\rho
(y)=k_D|y|$ correspond to global \cite{Greg00} and local \cite{Gher00}
strings respectively. In the case of more extra dimensions generalizations
of these models with $\Sigma =S^{D_2}$ were considered in \cite%
{Olas00,Gher00b,Oda01,Roes02}. The general case of a symmetric homogeneous
space $\Sigma $ with topologically non-trivial Yang-Mills fields is
investigated in \cite{Rand00,Rand00b}. In particular, when the bulk
cosmological constant is negative, a simple solution exists with $\sigma
(y)=\rho (y)=k_Dy$. For the case of $\Sigma $ with nonzero Ricci scalar one
can have a regular solution in the bulk with $\rho (y)={\mathrm{const}}$ and
Randall-Sundrum type warp in the Minkowski direction. The phenomenology of
this type of scenario has been considered in \cite{Davo03}.

Our main interest in this paper will be the Wightman function and the vacuum
expectation value (VEV) of the field square brought about by the presence of
two parallel infinite branes of codimension one, located at $y=a$ and $y=b$,
$a<b$. We will assume that on this boundaries the scalar field obeys the
mixed boundary conditions
\begin{equation}  \label{boundcond}
\left( \tilde A_y+\tilde B_y\partial _y\right) \varphi (x)=0,\quad y=a,b,
\end{equation}
with constant coefficients $\tilde A_y$, $\tilde B_y$. In the orbifolded
version of the model which corresponds to a higher dimensional
Randall-Sundrum braneworld these coefficients are expressed in terms of the
surface mass parameters and the curvature coupling of the scalar field (see
below). The imposition of boundary conditions on the quantum field modifies
the spectrum for the zero--point fluctuations and leads to the modification
of the VEVs for physical quantities compared with the case without
boundaries. The resulting quantum effects can either stabilize or
destabilize the branewolds and have to be taken into account in the
self-consistent formulation of the braneworld dynamics. As a first stage in
the investigations of local quantum effects, in this section we will
consider the positive frequency Wightman function defined as the expectation
value
\begin{equation}  \label{Wightfunc}
G^{+}(x,x^{\prime})=\langle 0|\varphi (x)\varphi (x^{\prime})|0\rangle ,
\end{equation}
where $|0\rangle $ is the amplitude for the vacuum state. The VEVs of the
field square and the energy-momentum tensor can be obtained from the
Wightman function in the coincidence limit with an additional
renormalization procedure. Instead of the Wightman function we could take
any other two-point function. The reason for our choice of the Wightman
function is related to the fact that this function also determines the
response of particle detectors in a given state of motion. Let $\varphi
_{\alpha }(x)$ be a complete set of positive frequency solutions to the
field equation (\ref{fieldeq}) satisfying boundary conditions (\ref%
{boundcond}) and $\alpha $ denotes a set of quantum numbers specifying the
solution. Expanding the field operator over this set of eigenfunctions and
using the commutation relations, the Wightman function is presented as the
mode sum:
\begin{equation}
G^{+}(x,x^{\prime })=\sum_{\alpha }\varphi _{\alpha }(x)\varphi _{\alpha
}^{\ast }(x^{\prime }).  \label{Wightvev}
\end{equation}

The symmetry of the bulk and boundary geometries under consideration allows
to present the corresponding eigenfunctions in the decomposed form
\begin{equation}
\varphi _{\alpha }(x^{M})=\phi _{\mathbf{k}}(x^{\mu })\psi _{\beta
}(X)f_{n}(y),  \label{eigfunc1}
\end{equation}%
with the standard Minkowskian modes in $R^{(D_1-1,1)}$:
\begin{equation}
\begin{split}
\phi _{\mathbf{k}}(x^{\mu })=&\frac{e^{-i\eta _{\mu \nu }k^{\mu }x^{\nu }}}{%
\sqrt{2\omega (2\pi )^{D_{1}-1}}},\quad k^{\mu }=(\omega ,\mathbf{k}), \\
\omega =&\sqrt{k^{2}+m_{\beta ,n}^{2}},\quad k=|\mathbf{k}|.
\label{branefunc1}
\end{split}%
\end{equation}%
The separation constants $m_{\beta ,n}$ are determined by the geometry of
the internal space $\Sigma $ and by the boundary conditions imposed on the
branes and will be given below. The modes $\psi _{\beta }(X)$ are
eigenfunctions for the operator $\Delta _{(\gamma )}+\zeta R_{(\gamma )}$:
\begin{equation}
\left[ \Delta _{(\gamma )}+\zeta R_{(\gamma )}\right] \psi _{\beta
}(X)=-\lambda _{\beta }^{2}\psi _{\beta }(X),  \label{eqint1}
\end{equation}%
with eigenvalues $\lambda _{\beta }^{2}$ and normalization condition
\begin{equation}  \label{normpsibet}
\int d^{D_2}X\, \sqrt{\gamma }\psi _{\beta }(X) \psi _{\beta ^{\prime
}}^{\ast }(X)=\delta _{\beta \beta ^{\prime }} ,
\end{equation}
$\Delta _{(\gamma )}$ is the Laplace-Beltrami operator for the metric $%
\gamma _{ij}$. In the consideration below we will assume that $\lambda
_{\beta }\geq 0$. Substituting eigenfunctions (\ref{eigfunc1}) into the
field equation (\ref{fieldeq}) and using equation (\ref{eqint1}), for the
function $f_{n}(y)$ one obtains the following equation%
\begin{equation}
\left[ -e^{\tau }\partial _{y}e^{-\tau }\partial _{y}+m^{2}-\zeta
F(y)+\lambda _{\beta }^{2}e^{2\rho }\right] f_{n}=m_{\beta ,n}^{2}e^{2\sigma
}f_{n} .  \label{eqforfn}
\end{equation}%
Equation (\ref{eqforfn}) is valid for any warp factors $\sigma (y)$ and $%
\rho (y)$. In this paper we consider the case of two equal warp factors,
with
\begin{equation}
\rho (y)=\sigma (y)=k_{D}y.  \label{rosigma}
\end{equation}%
In the case when $\Sigma $ is a torus, this corresponds to a toroidal
compactification of $AdS_{D+1}$ spacetime. For warp factors (\ref{rosigma})
equation (\ref{eqforfn}) takes the form%
\begin{equation}
\left[ -e^{D\sigma }\partial _{y}e^{-D\sigma }\partial _{y}+m^{2}-\zeta
D(D+1)k_{D}^{2}\right] f_{n}=\left( m_{\beta ,n}^{2}-\lambda _{\beta
}^{2}\right) e^{2\sigma } f_{n}.  \label{eqforfn1}
\end{equation}%
The operator in the left hand side does not depend on the internal quantum
numbers $\beta $. Thus in this case the eigenfunctions $f_{n} $ and the
combination $m_{n}^{2}=m_{\beta ,n}^{2}-\lambda _{\beta }^{2} $ do not
depend on $\beta $. As a result, the dependence on $\beta $ and $n$ of the
masses is factorized (see also \cite{Flac03b}):%
\begin{equation}
m_{\beta ,n}^{2}=m_{n}^{2}+\lambda _{\beta }^{2}.  \label{mbeta2}
\end{equation}%
For the region between the branes, $a<y<b$, the solution to equation (\ref%
{eqforfn1}) satisfying the boundary condition on the brane $y=a$ is given by
expression
\begin{equation}
f_{n}(y)=C_{n}e^{D\sigma /2}g_{\nu }(m_{n}z_{a},m_{n}z),  \label{fny}
\end{equation}%
where
\begin{equation}
g_{\nu }(u,v)=J_{\nu }(v)\bar{Y}_{\nu }^{(a)}(u)-\bar{J}_{\nu
}^{(a)}(u)Y_{\nu }(v),  \label{genu}
\end{equation}%
$J_{\nu }(x)$, $Y_{\nu }(x)$ are the Bessel and Neumann functions,
\begin{equation}  \label{zy}
z=\frac{e^{k_{D}y}}{k_{D}}, \quad z_{j}=\frac{e^{k_{D}j}}{k_{D}},\quad
j=a,b\, ,
\end{equation}
and
\begin{equation}
\nu =\sqrt{(D/2)^{2}-D(D+1)\zeta +m^{2}/k_{D}^{2}}.  \label{nu}
\end{equation}%
In formula (\ref{genu}) for a given function $F(x)$ we use the notation
\begin{equation}
\bar{F}^{(j)}(x)=A_{j}F(x)+B_{j}xF^{\prime }(x),\quad j=a,b,  \label{notbar}
\end{equation}%
with the coefficients
\begin{equation}
A_{j}=\tilde{A}_{j}+ \tilde{B}_{j}k_{D}D/2,\quad B_{j}=\tilde{B}_{j}k_{D}.
\label{AjBj}
\end{equation}
In the discussion below we will assume values of the curvature coupling
parameter for which $\nu $ is real. For imaginary $\nu $ the ground state
becomes unstable \cite{Brei82}. For a conformally coupled massless scalar
one has $\nu =1/2$ and the cylinder functions in Eq. (\ref{genu}) are
expressed via the elementary functions. For a minimally coupled massless
scalar $\nu =D/2$ and the same is the case in odd spatial dimensions. Note
that in terms of the coordinate $z$ introduced by relation (\ref{zy}), the
metric tensor from (\ref{metric}) with warp factors (\ref{rosigma}) is
conformally related to the metric of the direct product space $%
R^{(D_1,1)}\times \Sigma $ by the conformal factor $(k_D z)^{-2}$.

From the boundary condition on the brane $y=b$ we receive that the
eigenvalues $m_{n}$ have to be solutions to the equation
\begin{equation}
C_{\nu }^{ab}(z_{b}/z_{a},m_{n}z_{a})\equiv \bar{J}_{\nu }^{(a)}(m_{n}z_{a})%
\bar{Y}_{\nu }^{(b)}(m_{n}z_{b})-\bar{Y}_{\nu }^{(a)}(m_{n}z_{a})\bar{J}%
_{\nu }^{(b)}(m_{n}z_{b})=0.  \label{cnu}
\end{equation}%
This equation determines the tower of radial Kaluza-Klein (KK) masses. We
denote by $z=\gamma _{\nu ,n}$, $n=1,2,\ldots $, the zeros of the function $%
C_{\nu }^{ab}(\eta ,z)$ in the right half-plane of the complex variable $z$,
arranged in the ascending order, $\gamma _{\nu ,n}<\gamma _{\nu ,n+1}$. The
eigenvalues for $m_{n}$ are related to these zeros as
\begin{equation}
m_{n}=k_{D}\gamma _{\nu ,n}e^{-k_D a}=\gamma _{\nu ,n}/z_{a}.
\label{mntogam}
\end{equation}%
From the orthonormality condition for the functions $f_n(y)$ for the
coefficient $C_{n}$ in Eq. (\ref{fny}) one finds
\begin{equation}
C_{n}^{2}=\frac{\pi ^{2}\gamma _{\nu ,n}}{2k_{D}z_{a}^{2}}T_{\nu
}^{ab}\left( \eta ,\gamma _{\nu ,n}\right) ,\quad \eta =\frac{z_{b}}{z_{a}},
\label{cn}
\end{equation}%
where we have introduced the notation
\begin{equation}
T_{\nu }^{ab}(\eta ,u)=u\left\{ \frac{\bar{J}_{\nu }^{(a)2}(u)}{\bar{J}_{\nu
}^{(b)2}(\eta u)}\left[ A_{b}^{2}+B_{b}^{2}(\eta ^{2}u^{2}-\nu ^{2})\right]
-A_{a}^{2}+B_{a}^{2}(u^{2}-\nu ^{2})\right\} ^{-1}.  \label{Tnu}
\end{equation}%
Note that, as we consider the quantization in the region between the branes,
$z_{a}\leq z\leq z_{b}$, the modes defined by (\ref{fny}) are normalizable
for all real values of $\nu $ from Eq. (\ref{nu}).

Substituting the eigenfunctions (\ref{eigfunc1}) into the mode sum (\ref%
{Wightvev}), for the expectation value of the field product one finds%
\begin{eqnarray}
\langle 0|\varphi (x)\varphi (x^{\prime })|0\rangle &=&\frac{%
k_{D}^{D-1}(zz^{\prime })^{\frac{D}{2}}}{2^{D_{1}+1}\pi ^{D_{1}-3}z_{a}^{2}}%
\sum_{\beta }\psi _{\beta }(X)\psi _{\beta }^{\ast }(X^{\prime })  \notag \\
&\times &\int d\mathbf{k}\,e^{i\mathbf{k}(\mathbf{x}-\mathbf{x}^{\prime
})}\sum_{n=1}^{\infty }h_{\beta \nu }(\gamma _{\nu ,n})T_{\nu }^{ab}(\eta
,\gamma _{\nu ,n}),  \label{W11}
\end{eqnarray}%
where $\mathbf{x}=(x^{1},x^{2},\ldots ,x^{D_{1}-1})$ represents the spatial
coordinates in $R^{(D_1-1,1)}$, and%
\begin{equation}
h_{\beta \nu }(u)=ug_{\nu }(u,uz/z_{a})g_{\nu }(u,uz^{\prime }/z_{a})\frac{%
e^{i\sqrt{u^{2}/z_{a}^{2}+k^{2}+\lambda _{\beta }^{2}}}}{\sqrt{%
u^{2}/z_{a}^{2}+k^{2}+\lambda _{\beta }^{2}}}.  \label{hab}
\end{equation}%
As the expressions for the zeros $\gamma _{\nu ,n}$ are not explicitly
known, the form (\ref{W11}) of the Wightman function is inconvenient. In
addition, the terms in the sum over $n$ are highly oscillatory functions for
large values $n$. It is possible to overcome both these difficulties by
applying to the sum over $n$ the summation formula derived in Refs. \cite%
{Sahreview,Sahsph} by making use of the generalized Abel-Plana formula. For
a function $h(u)$ analytic in the right half-plane $\mathrm{Re}\,u>0$ this
formula has the form
\begin{eqnarray}
\frac{\pi ^{2}}{2}\sum_{n=1}^{\infty }h(\gamma _{\nu ,n})T_{\nu }(\eta
,\gamma _{\nu ,n}) &=&\int_{0}^{\infty }\frac{h(x)dx}{\bar{J}_{\nu
}^{(a)2}(x)+\bar{Y}_{\nu }^{(a)2}(x)}  \notag \\
&-& \frac{\pi }{4}\int_{0}^{\infty }dx\, \Omega _{a\nu }(x,\eta x)\left[
h(xe^{\frac{\pi i}{2}})+h(xe^{-\frac{\pi i}{2}})\right] ,  \label{cor3form}
\end{eqnarray}%
where
\begin{equation}
\Omega _{a\nu }(u,v) =\frac{\bar{K}_{\nu }^{(b)}(v)/\bar{K}_{\nu }^{(a)}(u)}{%
\bar{K}_{\nu }^{(a)}(u)\bar{I}_{\nu }^{(b)}(v)-\bar{K}_{\nu }^{(b)}(v)\bar{I}%
_{\nu }^{(a)}(u)},  \label{Omnu}
\end{equation}
$I_{\nu }(u)$ and $K_{\nu }(u)$ are the Bessel modified functions. Formula (%
\ref{cor3form}) is valid for functions $h(u)$ satisfying the conditions
\begin{equation}
|h(u)|<\varepsilon _{1}(x)e^{c_{1}|y|}\quad |u|\rightarrow \infty ,\quad
u=x+iy,  \label{cond31}
\end{equation}%
and $h(u)=o(u^{-1})$, $u\to 0$, where $c_{1}<2(\eta -1)$, $x^{2\delta
_{B_{a}0}-1}\varepsilon _{1}(x)\rightarrow 0$ for $x\rightarrow +\infty $.
Using the asymptotic formulae for the Bessel functions for large arguments
when $\nu $ is fixed (see, e.g., \cite{abramowiz}), we can see that for the
function $h_{\beta \nu }(u)$ from Eq. (\ref{hab}) the condition (\ref{cond31}%
) is satisfied if $z+z^{\prime }+|t-t^{\prime }|<2z_{b}$. In particular,
this is the case in the coincidence limit $t=t^{\prime }$ for the region
under consideration, $z_{a}<z,z^{\prime }<z_{b}$. Applying to the sum over $n
$ in Eq. (\ref{W11}) formula (\ref{cor3form}), one obtains
\begin{eqnarray}
\langle 0|\varphi (x)\varphi (x^{\prime })|0\rangle &=&\frac{%
k_{D}^{D-1}(zz^{\prime })^{D/2}}{2^{D_{1}}\pi ^{D_{1}-1}}\sum_{\beta }\psi
_{\beta }(X)\psi _{\beta }^{\ast }(X^{\prime })\int d\mathbf{k}\,e^{i\mathbf{%
k}(\mathbf{x}-\mathbf{x}^{\prime })}  \notag \\
&\times & \bigg\{ \frac{1}{z_{a}^{2}}\int_{0}^{\infty }\frac{h_{\beta \nu
}(u)du}{\bar{J}_{\nu }^{(a)2}(u)+\bar{Y}_{\nu }^{(a)2}(u)}-\frac{2}{\pi }%
\int_{\sqrt{k^{2}+\lambda _{\beta }^{2}}}^{\infty }du\,u\frac{\Omega _{a\nu
}(uz_{a},uz_{b})}{\sqrt{u^{2}-k^{2}-\lambda _{\beta }^{2}}}  \notag \\
&\times & G_{\nu }^{(a)}(uz_{a},uz)G_{\nu }^{(a)}(uz_{a},uz^{\prime })\cosh
\left[ \sqrt{u^{2}-k^{2}-\lambda _{\beta }^{2}}(t-t^{\prime })\right] %
\bigg\} .  \label{W13}
\end{eqnarray}%
where we have introduced notations
\begin{equation}
G_{\nu }^{(j)}(u,v) =I_{\nu }(v)\bar{K}_{\nu }^{(j)}(u)-\bar{I}_{\nu
}^{(j)}(u)K_{\nu }(v),\;j=a,b,  \label{Geab}
\end{equation}%
(the function with $j=b$ will be used below). Note that we have assumed
values of the coefficients $A_{j }$ and $B_{j }$ for which all zeros for Eq.
(\ref{cnu}) are real and have omitted the residue terms in the original
formula in Refs. \cite{Sahreview,Sahsph}. In the following we will consider
this case only.

By the way similar to that used in Ref. \cite{Saha04a}, it can be seen that
under the condition $z+z^{\prime }-|t-t^{\prime }|>2z_{a}$, the first term
in the figure braces in Eq. (\ref{W13}) is presented in the form
\begin{eqnarray}
\frac{1}{z_{a}^{2}}\int_{0}^{\infty }\frac{h_{\beta \nu }(u)du}{\bar{J}_{\nu
}^{(a)2}(u)+\bar{Y} _{\nu }^{(a)2}(u)} &=&\int_{0}^{\infty }duu\frac{e^{i%
\sqrt{u^{2}+k^{2}+\lambda _{\beta }^{2}}(t^{\prime }-t)}}{\sqrt{%
x^{2}+k^{2}+\lambda _{\beta }^{2}}}J_{\nu }(uz)J_{\nu }(uz^{\prime })  \notag
\\
&-&\frac{2}{\pi }\int_{\sqrt{k^{2}+\lambda _{\beta }^{2}}}^{\infty }du\,\ u%
\frac{\bar{I}_{\nu }^{(a)}(uz_{a})}{\bar{K}_{\nu }^{(a)}(uz_{a})}\frac{%
K_{\nu }(uz)K_{\nu }(uz^{\prime })}{\sqrt{u^{2}-k^{2}-\lambda _{\beta }^{2}}}
\notag \\
&\times & \cosh \!\left[ \sqrt{u^{2}-k^{2}-\lambda _{\beta }^{2}}%
(t-t^{\prime }) \right] .  \label{rel1term}
\end{eqnarray}%
Substituting this into formula (\ref{W13}), for the Wightman function one
finds
\begin{eqnarray}
\langle 0|\varphi (x)\varphi (x^{\prime })|0\rangle &=&\langle \varphi
(x)\varphi (x^{\prime })\rangle ^{(0)} +\langle \varphi (x)\varphi
(x^{\prime })\rangle ^{(a)}  \notag \\
&-& \frac{k_{D}^{D-1}(zz^{\prime })^{D/2}}{2^{D_{1}-1}\pi ^{D_{1}}}%
\sum_{\beta }\psi _{\beta }(X)\psi _{\beta }^{\ast }(X^{\prime })\int d%
\mathbf{k}\,e^{i\mathbf{k}(\mathbf{x}-\mathbf{x}^{\prime })}  \notag \\
&\times & \int_{\sqrt{k^{2}+\lambda _{\beta }^{2}}}^{\infty }du u G_{\nu
}^{(a)}(uz_{a},uz)G_{\nu }^{(a)}(uz_{a},uz^{\prime })  \notag \\
&\times & \frac{\Omega _{a\nu }(uz_{a},uz_{b})}{\sqrt{u^{2}-k^{2}-\lambda
_{\beta }^{2}}} \cosh \left[ \sqrt{u^{2}-k^{2}-\lambda _{\beta }^{2}}%
(t-t^{\prime })\right] .  \label{W15}
\end{eqnarray}%
Here the term
\begin{eqnarray}
\langle \varphi (x)\varphi (x^{\prime })\rangle ^{(0)}&=&\frac{%
k_{D}^{D-1}(zz^{\prime })^{D/2}}{2^{D_1}\pi ^{D_1-1}}\sum_{\beta }\psi
_{\beta }(X)\psi _{\beta }^{\ast }(X^{\prime })\int d\mathbf{k}\,e^{i\mathbf{%
k}(\mathbf{x}-\mathbf{x}^{\prime })}  \notag \\
&\times & \int_{0}^{\infty }du\,u\frac{e^{i\sqrt{u^{2}+k^{2}+\lambda _{\beta
}^{2}}(t^{\prime }-t)}}{\sqrt{u^{2}+k^{2}+\lambda _{\beta }^{2}}}J_{\nu }(u
z)J_{\nu }(u z^{\prime })  \label{WAdS}
\end{eqnarray}%
does not depend on the boundary conditions and is the Wightman function for
the $AdS_{D_1+1}\times \Sigma $ spacetime without branes. The second term on
the right of Eq. (\ref{W15}),
\begin{eqnarray}
\langle \varphi (x)\varphi (x^{\prime })\rangle ^{(a)} &=&-\frac{%
k_{D}^{D-1}(zz^{\prime })^{D/2}}{2^{D_{1}-1}\pi ^{D_{1}}}\sum_{\beta }\psi
_{\beta }(X)\psi _{\beta }^{\ast }(X^{\prime })\int d\mathbf{k}\,e^{i\mathbf{%
k}(\mathbf{x}-\mathbf{x}^{\prime })}  \notag \\
&\times &\int_{\sqrt{k^{2}+\lambda _{\beta }^{2}}}^{\infty }duu\frac{\bar{I}%
_{\nu }^{(a)}(uz_{a})}{\bar{K}_{\nu }^{(a)}(uz_{a})}\frac{K_{\nu }(uz)K_{\nu
}(uz^{\prime })}{\sqrt{u^{2}-k^{2}-\lambda _{\beta }^{2}}}  \notag \\
&\times & \cosh \!\left[ \sqrt{u^{2}-k^{2}-\lambda _{\beta }^{2}}%
(t-t^{\prime })\right] ,  \label{1bounda}
\end{eqnarray}%
does not depend on the parameters of the brane at $z=z_{b}$ and is induced
in the region $z>z_{a}$ by a single brane at $z=z_{a}$ when the boundary $%
z=z_{b}$ is absent. Thus the last term on the right of formula (\ref{W15})
is the part in the Wightman function induced by the presence of the second
brane. Hence, the application of the summation formula based on the
generalized Abel-Plana formula allowed us (i) to escape the necessity to
know the explicit expressions for the zeros $\gamma _{\nu ,n}$, (ii) to
extract from the two-point function the boundary-free and single brane
parts, (iii) to present the remained part in terms of integrals with the
exponential convergence in the coincidence limit. By the same way described
above for the Wightman function, any other two-point function can be
evaluated. Note that expression (\ref{WAdS}) for the boundary-free Wightman
function can also be written in the form
\begin{eqnarray}
\langle \varphi (x)\varphi (x^{\prime })\rangle ^{(0)} &=& k_D^{D-1}(z
z^{\prime})^{D/2} \sum_{\beta }\psi _{\beta }(X)\psi _{\beta }^{\ast
}(X^{\prime })  \notag \\
&& \times \int_{0}^{\infty }du \, u G_{R^{(D_1-1,1)}}^{+}(x^\mu , {x^{\prime}%
}^\mu ; \sqrt{u^2+\lambda ^2_\beta })J_{\nu }(m z)J_{\nu }(m z^{\prime }),
\label{WAdS1}
\end{eqnarray}
where $G_{R^{(D_1-1,1)}}^{+}(x^\mu , {x^{\prime}}^\mu ;m)$ is the Wightman
function for a scalar field with mass $m$ in the $D_1$-dimensional Minkowski
spacetime.

By using the identity
\begin{eqnarray}
&&\frac{\bar{K}_{\nu }^{(b)}(uz_{b})}{\bar{I}_{\nu }^{(b)}(uz_{b})}I_{\nu
}(uz)I_{\nu }(uz^{\prime })-\frac{\bar{I}_{\nu }^{(a)}(uz_{a})}{\bar{K}_{\nu
}^{(a)}(uz_{a})}K_{\nu }(uz)K_{\nu }(uz^{\prime })=  \notag \\
&&\sum_{j=a,b}n^{(j)}\Omega _{j\nu }(uz_{a},uz_{b})G_{\nu
}^{(j)}(uz_{j},uz)G_{\nu }^{(j)}(uz_{j},uz^{\prime }),  \label{ident11}
\end{eqnarray}%
where $n^{(j)}=1$ for the region $y>j$ and $n^{(j)}=-1$ for the region $y<j$
(hence, in (\ref{ident11}) one has $n^{(a)}=1$, $n^{(b)}=-1$ as we consider
the region $a<y<b$), and
\begin{equation}
\Omega _{b\nu }(u,v)=\frac{\bar{I}_{\nu }^{(a)}(u)/\bar{I}_{\nu }^{(b)}(v)}{%
\bar{K}_{\nu }^{(a)}(u)\bar{I}_{\nu }^{(b)}(v)-\bar{K}_{\nu }^{(b)}(v)\bar{I}%
_{\nu }^{(a)}(u)},  \label{Omnub}
\end{equation}%
the Wightman function in the region $z_{a}\leq z\leq z_{b}$ can also be
presented in the form
\begin{eqnarray}
\langle 0|\varphi (x)\varphi (x^{\prime })|0\rangle  &=&\langle \varphi
(x)\varphi (x^{\prime })\rangle ^{(0)}+\langle \varphi (x)\varphi (x^{\prime
})\rangle ^{(b)}  \notag \\
&-&\frac{k_{D}^{D-1}(zz^{\prime })^{D/2}}{2^{D_{1}-1}\pi ^{D_{1}}}%
\sum_{\beta }\psi _{\beta }(X)\psi _{\beta }^{\ast }(X^{\prime })\int d%
\mathbf{k}\,e^{i\mathbf{k}(\mathbf{x}-\mathbf{x}^{\prime })}  \notag \\
&\times &\int_{\sqrt{k^{2}+\lambda _{\beta }^{2}}}^{\infty }duu\frac{\Omega
_{b\nu }(uz_{a},uz_{b})}{\sqrt{u^{2}-k^{2}-\lambda _{\beta }^{2}}}  \notag \\
&\times &G_{\nu }^{(b)}(uz_{b},uz)G_{\nu }^{(b)}(uz_{b},uz^{\prime })\cosh
\left[ \sqrt{u^{2}-k^{2}-\lambda _{\beta }^{2}}(t-t^{\prime })\right] .
\label{W17}
\end{eqnarray}%
In this formula
\begin{eqnarray}
\langle \varphi (x)\varphi (x^{\prime })\rangle ^{(b)} &=&-\frac{%
k_{D}^{D-1}(zz^{\prime })^{D/2}}{2^{D_{1}-1}\pi ^{D_{1}}}\sum_{\beta }\psi
_{\beta }(X)\psi _{\beta }^{\ast }(X^{\prime })\int d\mathbf{k}\,e^{i\mathbf{%
k}(\mathbf{x}-\mathbf{x}^{\prime })}  \notag \\
&\times &\int_{\sqrt{k^{2}+\lambda _{\beta }^{2}}}^{\infty }duu\frac{\bar{K}%
_{\nu }^{(b)}(uz_{b})}{\bar{I}_{\nu }^{(b)}(uz_{b})}\frac{I_{\nu }(uz)I_{\nu
}(uz^{\prime })}{\sqrt{u^{2}-k^{2}-\lambda _{\beta }^{2}}}  \notag \\
&\times &\cosh \!\left[ \sqrt{u^{2}-k^{2}-\lambda _{\beta }^{2}}(t-t^{\prime
})\right]   \label{1boundb}
\end{eqnarray}%
is the boundary part induced in the region $z<z_{b}$ by a single brane at $%
z=z_{b}$ when the brane $z=z_{a}$ is absent. Note that in the formulae given
above the integration over angular part can be done by using the formula
\begin{equation}
\int d\mathbf{k}\,\frac{e^{i\mathbf{k}\mathbf{x}}F(k)}{(2\pi )^{\frac{D_{1}-1%
}{2}}}=\int_{0}^{\infty }dk\,k^{D_{1}-2}F(k)\frac{J_{(D_{1}-3)/2}(k|\mathbf{x%
}|)}{(k|\mathbf{x}|)^{\frac{D_{1}-3}{2}}},  \label{angint}
\end{equation}%
for a given function $F(k)$. Combining two forms, formulae (\ref{W15}) and (%
\ref{W17}), we see that the expressions for the Wightman function in the
region $z_{a}\leq z\leq z_{b}$ is symmetric under the interchange $%
a\rightleftarrows b$ and $I_{\nu }\rightleftarrows K_{\nu }$. Note that the
expression for the Wightman function is not symmetric with respect to the
interchange of the brane indices. The reason for this is that the boundaries
have nonzero extrinsic curvature tensors and two sides of the boundaries are
not equivalent. In particular, for the geometry of a single brane the VEVs
are different for the regions on the left and on the right of the brane. For
the case $\Sigma =R^{D_{2}}$ one has $\psi _{\beta }(X)=(2\pi
)^{-D_{2}/2}e^{i{\mathbf{KX}}}$, where $\beta ={\mathbf{K}}$ is the $D_{2}$%
-dimensional wave vector and replacing in the formulae above $\sum_{\beta
}\rightarrow (2\pi )^{-D_{2}}\int d{\mathbf{K}}$ we obtain the results for
the $AdS_{D+1}$ bulk investigated in Ref. \cite{Saha04a}. Examining (\ref%
{1boundb}) tells one that the brane induced part in the Wightman
function vanishes if one or both its arguments lie on the AdS
boundary corresponding to $z=0$. Note that this does not mean that
the presence of the brane in the bulk has no consequences in the
conformal field theory on the boundary, translated from the bulk
physics by using the AdS/CFT correspondence. Here one should bear
in mind that in the AdS/CFT correspondence the asymptotic behavior
of the bulk modes $\varphi _{\alpha }(x^{\mu },X^{i},z)$ is
related to the corresponding boundary fields $\varphi _{0\alpha
}(x^{\mu },X^{i})$ by the relation $\varphi _{0\alpha }(x^{\mu
},X^{i})\sim \lim_{z\rightarrow 0}z^{-D/2-\nu }\varphi _{\alpha
}(x^{\mu },X^{i},z)$. It follows from here that the presence of
the brane in the bulk leads to the additional contribution in the
boundary two point function which is obtained from (\ref{1boundb})
as the limit $\lim_{z,z'\rightarrow 0}(zz^{\prime })^{-D/2-\nu
}\langle \varphi (x)\varphi (x^{\prime })\rangle ^{(b)}$. This
limit is nonzero and gives nontrivial consequences in the boundary
conformal field theory. It would be interesting to discuss the
holographic interpretation of the brane degrees of freedom in
terms of conformal field theory. However, such a discussion is
beyond the scope of the present paper.

Now let us consider the application of the results given above to the higher
dimensional generalization of the Randall-Sundrum braneworld based on the
bulk $AdS_{D_1+1}\times \Sigma $. In the corresponding model $y$ coordinate
is compactified on an orbifold, $S^1/Z_2$ of length $l$, with $-l\leq y\leq l
$. The orbifold fixed points at $y=0$ and $y=l$ are the locations of two $D$%
-dimensional branes. The corresponding line element has the form (\ref%
{metric}) with $\sigma (y)=\rho (y)=k_D|y|$. The absolute value sign here
leads to $\delta $-type contributions to the corresponding Ricci scalar,
located on the branes. As a result $\delta $-terms appear in the equation
for the eigenfunction $f_n(y)$. Additional $\delta $-terms come from the
surface action of the scalar field with mass parameters $c_1$ and $c_2$ on
the branes at $y=0$ and $y=l$ respectively. The $\delta $-terms in the
equation for the function $f_n(y)$ in combination with the $Z_2$ symmetry
lead to the boundary conditions on this function. By the way similar to that
for the usual Randall-Sundrum model (see, for instance, \cite%
{Gher00,Flac01b,Saha04a}) it can be seen that for untwisted scalar the
boundary conditions are Robin type with the coefficients
\begin{equation}  \label{AtildeRS}
\frac{\tilde A_a}{\tilde B_a} = -\frac{1}{2}(c_1+4D\zeta k_D),\quad \frac{%
\tilde A_b}{\tilde B_b} = -\frac{1}{2}(-c_2+4D\zeta k_D).
\end{equation}
For twisted scalar field Dirichlet boundary conditions are obtained. Note
that in the orbifolded version the integration in the normalization integral
for the function $f_n(y)$ goes over two copies of the bulk manifold. This
leads to the additional coefficient $1/2$ in the expression for the
normalization coefficient $C_n$ in (\ref{cn}). Hence, the Wightman function
in the higher dimensional Randall-Sundrum braneworld is given by formula (%
\ref{W15}) or equivalently by (\ref{W17}) with an additional factor 1/2 and
with Robin coefficients given by Eq. (\ref{AtildeRS}). The one-loop
effective potential and the problem of moduli stabilization in this model
with zero mass parameters $c_j$ are discussed in Ref. \cite{Flac03b}. In
particular, a scenario is proposed where supersymmetry is broken near the
fundamental Planck scale, and the hierarchy between the electroweak and
effective Planck scales is generated by a combination of redshift and large
volume effects. The masses of the corresponding KK modes along $\Sigma $ are
of the order of TeV. The phenomenology of this type braneworld models is
discussed in Refs. \cite{Rand00b,Chac00,Mult02,Flac03b}.

\section{Vacuum polarization by a single brane}

\label{sec:vevphi2}

In this section we will consider the VEV of the field square induced by a
single brane located at $y=a$. As it has been shown in the previous section
the Wightman function for this geometry is presented in the form
\begin{equation}
\langle 0|\varphi (x)\varphi (x^{\prime })|0\rangle =\langle \varphi
(x)\varphi (x^{\prime })\rangle ^{(0)}+\langle \varphi (x)\varphi (x^{\prime
})\rangle ^{(a)},  \label{Wigh1b}
\end{equation}%
where the boundary-free part on the left is defined by formula (\ref{WAdS}).
The brane induced part $\langle \varphi (x)\varphi (x^{\prime })\rangle
^{(a)} $ is given by formula (\ref{1bounda}) in the region $z>z_{a}$ and by
formula (\ref{1boundb}) with replacement $z_{b}\rightarrow z_{a}$ in the
region $z<z_{a}$. The VEV of the field square is obtained from the Wightman
function taking the coincidence limit of the arguments. Similar to Eq. (\ref%
{Wigh1b}) this VEV is presented as the sum of boundary-free and boundary
induced parts:
\begin{equation}  \label{phi21b}
\langle 0|\varphi ^2|0\rangle =\langle \varphi ^2\rangle ^{(0)}+\langle
\varphi ^2\rangle ^{(a)}.
\end{equation}
For the part without boundaries the direct substitution of the coincident
arguments in the expression of the Wightman function leads to the divergent
expression. For the regularization of this divergence the combination of the
zeta function method for the sum over $\beta $ and the dimensional
regularization method for the integral over $k$ can be used. By this way,
taking the coincidence limit in the expression (\ref{WAdS}) and evaluating
the ${\mathbf{k}}$-integral, for the VEV of the field square in the
boundary-free case one finds
\begin{equation}  \label{phi2bfree}
\langle \varphi ^2\rangle ^{(0)}=\frac{k_D^{D-1}z^D}{2^{D_1}\pi ^{\frac{D_1}{%
2}}}\Gamma \left( 1-\frac{D_1}{2}\right) \int_{0}^{\infty }du\, u \, \zeta
\left( 1-\frac{D_1}{2},X;u\right) J_{\nu }^2(xz),
\end{equation}
where we have introduced the local spectral zeta function
\begin{equation}  \label{zetalocal}
\zeta \left( s,X;m\right) =\sum_{\beta } \frac{|\psi _{\beta }(X)|^2}{%
(\lambda _{\beta }^2+m^2)^{s}}
\end{equation}
for the operator $\triangle _{(\gamma )}+\zeta R_{(\gamma )}-m^2$. This
corresponds to a scalar field with the mass $m$ propagating on background of
the manifold $\Sigma $. For $\Sigma =R^{D_2}$ from (\ref{phi2bfree}) we
obtain the standard result for the $AdS_{D+1}$ bulk. In this case the VEV of
the field square does not depend on the coordinate $z$. For the bulk $%
AdS\times \Sigma $, even in the case of homogeneous compact subspace $\Sigma
$, this VEV depends on $z$. We explicitly illustrate this below in section %
\ref{sec:example} for the simple case $\Sigma =S^1$.

For points away from the brane the boundary induced part in the Wightman
function is finite in the coincidence limit and we can directly put $%
x=x^{\prime }$. Introducing a new integration variable $v=\sqrt{%
u^{2}-k^{2}-\lambda _{\beta }^{2}}$, transforming to the polar coordinates
in the plane $(v,k)$ and integrating over angular part, the following
formula can be derived
\begin{equation}
\int_{0}^{\infty }dk\int_{\sqrt{k^{2}+\lambda _{\beta }^{2}}}^{\infty }\frac{%
k^{D_{1}-2}uf(u)du}{\sqrt{u^{2}-k^{2}-\lambda _{\beta }^{2}}}=\frac{\sqrt{%
\pi }\Gamma \left( \frac{D_{1}-1}{2}\right) }{2\Gamma \left( \frac{D_{1}}{2}%
\right) }\int_{0}^{\infty }duu^{D_{1}-1}f(\sqrt{u^{2}+\lambda _{\beta }^{2}}%
).  \label{rel3}
\end{equation}%
By using this formula and Eq. (\ref{1bounda}), the boundary induced VEV for
the field square in the region $z>z_{a}$ is presented in the form
\begin{equation}  \label{phi2beta}
\langle \varphi ^{2}\rangle ^{(a)}=\sum_{\beta }|\psi _{\beta }(X)|^2\langle
\varphi ^{2}\rangle ^{(a)}_{\beta },
\end{equation}
where the contribution of a given KK mode along $\Sigma $ is determined by
formula
\begin{equation}
\langle \varphi ^{2}\rangle ^{(a)}_{\beta } =-\frac{k_{D}^{D-1}z^{D}}{
2^{D_{1}-1}\pi ^{\frac{D_{1}}{2}}\Gamma \left( \frac{D_{1}}{2}\right) }
\int_{\lambda _\beta }^{\infty }du\,u (u^2- \lambda _\beta ^2)^{\frac{D_{1}}{%
2}-1}\frac{\bar{I}_{\nu }^{(a)}(uz_{a})}{\bar{K}_{\nu }^{(a)}(uz_{a})}K_{\nu
}^{2}(uz).  \label{phi2spl}
\end{equation}%
The corresponding formula in the region $z<z_{a}$ is obtained from Eq. (\ref%
{1boundb}) by a similar way and differs from Eq. (\ref{phi2spl}) by the
replacements $I_{\nu }\rightleftarrows K_{\nu }$. Using the properties of
the Bessel modified functions, it can be seen that $\langle \varphi
^{2}\rangle ^{(a)}<0$ for Dirichlet boundary condition ($B_a=0$) and $%
\langle \varphi ^{2}\rangle ^{(a)}>0$ for the ratio of the coefficients in
the interval $-\nu <A_a/B_a<\nu $ in both regions $z<z_a $ and $z>z_a$. In
the general case, the VEV may change the sign as a function of $z$. In the
models with a homogeneous internal space $\Sigma $, the sum over $\beta $ in
Eq. (\ref{phi2beta}) and, hence, the VEV of the field square do not depend
on the coordinates $X^{i}$ in this space.

For the comparison with the case of the bulk spacetime $AdS_{D_1+1}$ when
the internal space is absent, it is useful in addition to the VEV (\ref%
{phi2beta}) to consider the VEV integrated over the subspace~$\Sigma $:
\begin{equation}  \label{phi2integrated}
\langle \varphi ^{2}\rangle ^{(a)}_{{\mathrm{integrated}}}=\int_{\Sigma }
d^{D_2}X\sqrt{\gamma }\, \langle \varphi ^{2}\rangle ^{(a)} e^{-D_2k_Dy}
=e^{-D_2k_Dy}\sum_{\beta } \langle \varphi ^{2}\rangle ^{(a)}_{\beta }.
\end{equation}
Comparing this integrated VEV with the corresponding formula from Ref. \cite%
{Saha04a}, we see that the contribution of the zero KK mode ($\lambda
_{\beta }=0$) in Eq. (\ref{phi2integrated}) differs from the VEV of the
field square in the bulk $AdS_{D_1+1}$ by the order of the modified Bessel
functions: for the latter case $\nu \to \nu _1$ with $\nu _1$ defined by Eq.
(\ref{nu}) with the replacement $D\to D_1$. Note that for $\zeta \leq \zeta
_{D+D_1+1}$ one has $\nu \geq \nu _1$. In particular, this is the case for
minimally and conformally coupled scalar fields. If the internal space $%
\Sigma $ is a one-parameter manifold of size $L$ then one has $\lambda
_{\beta }\sim 1/L$. By taking into account that from the normalization
condition it follows that the function $\psi _{\beta }(X)$ contains the
factor $L^{-D_2/2}$, we see that $\langle \varphi ^{2}\rangle ^{(a)}$ is a
function on $z/z_a$ and $L/z_a$. The first ratio is related to the proper
distance from the brane by the equation
\begin{equation}  \label{propdist}
z/z_a=e^{k_D(y-a)}.
\end{equation}
Hence, for a given size $L$ of the internal space, the VEV of the field
square in addition to the proper distance from the brane, also depends on
the absolute position of the brane in the bulk. Note that in the case of AdS
bulk the corresponding quantity depends only on the proper distance from the
brane \cite{Saha04a}. To discuss the physics from the point of view of an
observer residing on the brane $y=a$, it is convenient to introduce rescaled
coordinates $x^{\prime \mu }=e^{-k_Da}x^{\mu }$. For this observer the
physical size of the subspace $\Sigma$ is $L_a=Le^{-k_Da}$ and the
corresponding KK masses are rescaled by the warp factor, $\lambda _{\beta
}^{(a)}=\lambda _{\beta }e^{k_Da}$. Now we see that the VEV induced by a
single brane is a function of the proper distance from the brane and on the
ratio $L_a/(1/k_D)$ of the physical size for the internal space $\Sigma $
for an observer residing on the brane to the AdS curvature radius.

As a partial check for derived formulae let us consider the limit $k_D\to 0$%
. This corresponds to a boundary on the bulk $R^{(D_1,1)}\times \Sigma $.
For $k_D\to 0$, from (\ref{nu}) we see that the order $\nu $ of the
cylindrical functions is large. Introducing in (\ref{phi2spl}) the new
integration variable $v=u/\nu $, we can replace the Bessel modified
functions by their uniform asymptotic expansions for large values of the
order. To the leading order this gives:
\begin{eqnarray}  \label{phi2splMink}
\langle \varphi ^{2}\rangle ^{(a)}& \approx & \langle \varphi ^{2}\rangle
^{(a)}_{R^{(D_1,1)}\times \Sigma }=-\frac{(4\pi )^{- \frac{D_1}{2}}}{\Gamma
\left( \frac{D_1}{2}\right)}\sum_{\beta }\left| \psi _{\beta }(X)\right| ^2
\notag \\
&& \times \int_{\sqrt{m^2+\lambda _{\beta }^{2}}}^{\infty } du\,
(u^2-m^2-\lambda _{\beta }^{2})^{\frac{D_1}{2}-1}\frac{e^{-2u|y-a|}}{\tilde
c_{a}(u)}.
\end{eqnarray}
Here and below we use the notations
\begin{equation}  \label{cj}
\tilde c_{j}(u)=\frac{\tilde A_j-n^{(j)}\tilde B_j u}{\tilde
A_j+n^{(j)}\tilde B_j u}, \quad j=a,b,
\end{equation}
and $n^{(j)}$ is defined after formula (\ref{ident11}).

For the points on the brane the VEV given by formula (\ref{phi2spl})
diverges due to the divergence of the $u$-integral in the upper limit. The
surface divergences in the renormalized VEVs of the local physical
observables result from the idealization of the boundaries as perfectly
smooth surfaces which are perfect reflectors at all frequencies, and are
well known in quantum field theory with boundaries (see, for instance, \cite%
{Deut79,Kenn80,Birrell}). It seems plausible that such effects as surface
roughness, or the microstructure of the boundary on small scales can
introduce a physical cutoff needed to produce finite values of surface
quantities. In brane models the imperfectness would come from the quantum
gravity effects on the Planck scale. An alternative mechanism for
introducing a cutoff which removes singular behavior on boundaries is to
allow the position of the boundary to undergo quantum fluctuations \cite%
{Ford98}. Such fluctuations smear out the contribution of the high frequency
modes without the need to introduce an explicit high frequency cutoff. In
order to find the renormalized value of the field square on the brane many
regularization techniques are available nowadays. In particular, the
generalized zeta function method is in general very powerful to give
physical meaning to the divergent quantities. This method has been used in
Ref. \cite{Saha04} for the evaluation of the field square on the brane in
the model with $AdS_{D+1}$ bulk. Another possibility is to remove from (\ref%
{phi2spl}) the terms diverging in the limit $z\to z_a$. The relation between
various procedures for the evaluation of the renormalized VEV of the field
square on the brane in the thin wall approximation is recently discussed in
Ref. \cite{Pujo05}. In particular, on the specific example it has been shown
that different methods give the same result up to finite renormalization of
the surface mass and extrinsic curvature counter-terms.

From the divergent behavior of the vacuum fluctuations on the brane it
follows that for the points near the brane the main contribution into the
integral comes from large values of $u$. Assuming $\lambda _{\beta
}|z-z_a|\ll 1$, $k_D|y-a|\ll 1$ and replacing the Bessel modified functions
by their asymptotic expansions for large values of the argument, to the
leading order for the contribution of the mode with a given $\beta $ one
finds
\begin{equation}  \label{phi2splnear}
\langle \varphi ^{2}\rangle ^{(a)}_{\beta }\approx -\frac{%
(k_Dz_a)^{D-1}\kappa (B_a)}{(4\pi )^{\frac{D_1+1}{2}}|z-z_a|^{D_1-1}}\Gamma
\left( \frac{D_1-1}{2}\right) ,
\end{equation}
where the notation
\begin{equation}  \label{kappa}
\kappa (B_a)=2\delta _{B_a0}-1
\end{equation}
is introduced. Note that in this limit the proper distance from the brane is
much smaller compared with the AdS curvature radius and the proper size $L_a$
of the internal manifold $\Sigma $.

Now let us consider the contribution of the large KK masses along the
internal subspace $\Sigma $, assuming that $z_{a}\lambda _{\beta },z\lambda
_{\beta } \gg 1$. This corresponds to the KK masses $\lambda _{\beta }^{(a)}$
much larger than the AdS energy scale $k_D$. In particular, for $L_a\ll 1/k_D
$ these conditions are satisfied for all modes along $\Sigma $ with nonzero
KK masses. Introducing a new integration variable $v=u/\lambda _{\beta }$,
we can replace the Bessel modified functions by their asymptotic expressions
for large values of the argument. For the contribution of the mode with a
given $\beta $ one finds
\begin{equation}  \label{phi2spllargelamb}
\langle \varphi ^{2}\rangle ^{(a)}_{\beta }\approx -\frac{(k_Dz)^{D- 1}}{%
(4\pi )^{\frac{D_1}{2}}\Gamma \left( \frac{D_1}{2}\right)} \int_{\lambda
_{\beta }}^{\infty } du\, (u^2-\lambda _{\beta }^{2})^{\frac{D_1}{2}-1}\frac{%
e^{-2u|z-z_a|}}{c_{a}(u z_a)}.
\end{equation}
where we have introduced the notation
\begin{equation}  \label{cj1}
c_{j}(u)=\frac{ A_j-n^{(j)}B_j u}{A_j+n^{(j)}B_j u}, \quad j=a,b,
\end{equation}
and $n^{(j)}$ is defined after formula (\ref{ident11}). Additionally
assuming that $z_a\lambda _{\beta }\gg A_a/B_a$ or $B_a=0$, one has
\begin{equation}  \label{phi2spllargelamb1}
\langle \varphi ^{2}\rangle ^{(a)}_{\beta }\approx -\frac{(k_Dz)^{D-1}\kappa
(B_a)}{2^{D_1}\pi ^{\frac{D_1+1}{2}}}\left| \frac{\lambda _{\beta }}{z-z_a}%
\right| ^{\frac{D_1-1}{2}} K_{\frac{D_1-1}{2}}(2\lambda _{\beta }|z-z_a|).
\end{equation}
From this formula it follows that for $\lambda _{\beta }|z-z_a|\gg 1$ (the
observation point is not too close to the brane) the contribution of KK
modes with large $\lambda _{\beta }$ is exponentially suppressed.

For small values $z_a$, $z_a\ll z,1/\lambda _{\beta }$, using the asymptotic
formulae for the Bessel modified functions for small values of the argument,
we obtain the formula
\begin{equation}  \label{phi2splsmallza}
\langle \varphi ^{2}\rangle ^{(a)}_{\beta }\approx -\frac{2^{2-D_1-2\nu
}k_D^{D-1}z^Dz_a^{2\nu }}{\pi ^{\frac{D_1}{2}}\Gamma \left( \frac{D_1}{2}%
\right) \nu \Gamma ^2(\nu )c_a(\nu )} \int_{\lambda _{\beta }}^{\infty }du\,
u^{2\nu +1}(u^2-\lambda _{\beta }^2)^{\frac{D_1}{2}-1}K_{\nu }^2(uz).
\end{equation}
This limit corresponds to large distances from the brane, $k_D(y-a)\gg 1$
and small KK masses $\lambda _{\beta }^{(a)}$ compared with the AdS energy
scale, $\lambda _{\beta }^{(a)}\ll k_D$. In particular, from Eq. (\ref%
{phi2splsmallza}) it follows that for fixed values $k_D$, $y$, $\lambda
_{\beta }$ the brane induced VEV in the region $z>z_a$ vanishes as $z_a^{2
\nu }$ when the brane position tends to the AdS boundary, $z_a\to 0$.
Formula (\ref{phi2splsmallza}) is further simplified for two subcases. In
the case $z_a\ll \lambda _{\beta }^{-1}\ll z$ or equivalently $%
k_De^{-k_D(y-a)}\ll \lambda _{\beta }^{(a)}\ll k_D$, the main contribution
into the integral comes from the lower limit ant to the leading order we
find
\begin{equation}  \label{phi2splsmallza1}
\langle \varphi ^{2}\rangle ^{(a)}_{\beta }\approx -\frac{%
k_D^{D-1}z^{D_2}(z_a\lambda _{\beta })^{2\nu }(z\lambda _{\beta })^{\frac{D_1%
}{2}-1}}{2^{D_1+2\nu }\pi ^{\frac{D_1}{2}-1}\nu \Gamma ^2(\nu )c_{a}(\nu )}
e^{-2z\lambda _{\beta }},
\end{equation}
with the exponential suppression of the boundary induced VEV. In the limit $%
z_a\ll z\ll \lambda _{\beta }^{-1}$ (small KK masses, $\lambda _{\beta
}^{(a)}\ll k_De^{-k_D(y-a)}$) to the leading order we can put 0 in the lower
limit of the integral. Then the integral is evaluated by the standard
formula for the integrals involving the square of the MacDonald function and
one finds (see, for instance, \cite{Prudnikov2})
\begin{equation}  \label{phi2splsmallza2}
\langle \varphi ^{2}\rangle ^{(a)}_{\beta }\approx -\frac{k_D^{D-1}z^{D_2}}{%
2^{D_1}\pi ^{\frac{D_1-1}{2}}c_{a}(\nu )} \frac{\Gamma \left( \frac{D_1}{2}%
+2\nu \right) \Gamma \left( \frac{D_1}{2}+\nu \right) }{\nu \Gamma ^2(\nu
)\Gamma \left( \frac{D_1+1}{2}+\nu \right)} \left( \frac{z_a}{2z}\right)
^{2\nu }.
\end{equation}
In particular, this formula is valid for the zero mode.

In the limit $z\ll z_{a},1/\lambda _{\beta }$, again using the asymptotics
of the modified Bessel functions for small values of the argument, we obtain
the formula
\begin{equation}
\langle \varphi ^{2}\rangle _{\beta }^{(a)}\approx -\frac{2^{2-D_{1}-2\nu
}k_{D}^{D-1}z^{D+2\nu }}{\pi ^{\frac{D_{1}}{2}}\Gamma \left( \frac{D_{1}}{2}%
\right) \Gamma ^{2}(\nu +1)}\int_{\lambda _{\beta }}^{\infty }du\,u^{2\nu
+1}(u^{2}-\lambda _{\beta }^{2})^{\frac{D_{1}}{2}-1}\frac{\bar{K}_{\nu
}^{(a)}(uz_{a})}{\bar{I}_{\nu }^{(a)}(uz_{a})}.  \label{phi2splsmallz}
\end{equation}%
This limit corresponds to large proper distances from the brane $%
k_{D}(a-y)\gg 1$ in the region $y<a$. As we see for fixed values $k_{D}$, $%
z_{a}$, $\lambda _{\beta }$ the brane induced VEV vanishes as $z^{D+2\nu }$
when the observation point tends to the AdS boundary. As it has been
explained in the previous section [see paragraph after formula (\ref{angint}%
)], this does not mean that the presence of the brane in the bulk does not
contribute to the corresponding VEV in the boundary conformal field theory
within the framework of AdS/CFT duality. To further simplify the formula (%
\ref{phi2splsmallz}) we consider two subcases. For the case $z\ll \lambda
_{\beta }^{-1}\ll z_{a}$ corresponding to $e^{k_{D}(y-a)}\ll k_{D}/\lambda
_{\beta }^{(a)}\ll 1$, to the leading order we obtain the following formula
\begin{equation}
\langle \varphi ^{2}\rangle _{\beta }^{(a)}\approx -\frac{%
k_{D}^{D-1}z^{D_{2}}c_{a}(z_{a}\lambda _{\beta })e^{-2z_{a}\lambda _{\beta }}%
}{2^{D_{1}+2\nu -1}\pi ^{\frac{D_{1}}{2}-1}\Gamma ^{2}(\nu +1)}(z\lambda
_{\beta })^{\frac{D_{1}}{2}+2\nu }\left( \frac{z}{z_{a}}\right) ^{\frac{D_{1}%
}{2}}.  \label{phi2splsmallz1}
\end{equation}%
In the opposite limit for $z_{a}$, $z\ll z_{a}\ll \lambda _{\beta }^{-1}$,
which corresponds to small KK masses, $\lambda _{\beta }^{(a)}\ll k_{D}$,
the lower limit in the integral can be replaced by 0 and to the leading
order the contribution of the mode with a given $\beta $ does not depend on $%
\lambda _{\beta }$:
\begin{equation}
\langle \varphi ^{2}\rangle _{\beta }^{(a)}\approx -\frac{2^{2-D_{1}-2\nu
}k_{D}^{D-1}z^{D_{2}}}{\pi ^{\frac{D_{1}}{2}}\Gamma \left( \frac{D_{1}}{2}%
\right) \Gamma ^{2}(\nu +1)}\left( \frac{z}{z_{a}}\right) ^{D_{1}+2\nu
}\int_{0}^{\infty }du\,u^{D_{1}+2\nu -1}\frac{\bar{K}_{\nu }^{(a)}(u)}{\bar{I%
}_{\nu }^{(a)}(u)}.  \label{phi2splsmallz2}
\end{equation}

Next we consider the limit $z\lambda _{\beta }\gg 1$ when $z_a\lambda
_{\beta }$ is fixed, corresponding to $e^{k_D(y-a)}\gg k_D/\lambda _{\beta
}^{(a)}$, $y>a$. The main contribution to the integral in Eq. (\ref{phi2spl}%
) comes from the lower limit and to the leading order one finds
\begin{equation}  \label{phi2spllargez}
\langle \varphi ^{2}\rangle ^{(a)}_{\beta }\approx -\frac{%
(k_Dz)^{D-1}\lambda _{\beta }^{\frac{D_1}{2}-1}}{2^{D_1+1}\pi ^{\frac{D_1}{2}%
-1}z^{\frac{D_1}{2}}} \frac{\bar{I}_{\nu }^{(a)}(z_a\lambda _{\beta })}{\bar{%
K}_{\nu }^{(a)}(z_a\lambda _{\beta })} e^{-2z\lambda _{\beta }},
\end{equation}
with the exponentially suppressed VEV. In particular, the contribution of
the nonzero KK modes along $\Sigma $ exponentially vanishes when the
observation point tends to the AdS horizon, $z\to \infty $. Note that for
the zero mode ($\lambda _{\beta }=0$) the brane induced VEV near the AdS
horizon behaves as $z^{D_2-2\nu }$ (see formula (\ref{phi2splsmallza2})). In
the purely AdS bulk ($D_2=0$) this VEV vanishes on the horizon for $\nu >0$.
For an internal spaces with $D_2>2\nu $ the VEV diverges on the horizon.
Note that for a conformally coupled massless scalar and $D_2=1$ the boundary
induced VEV takes nonzero finite value on the horizon.

In the limit when the brane position tends to the AdS horizon, $z_a\to
\infty $, for massive KK modes along $\Sigma $ the main contribution into
the VEV of the field square in the region $z<z_a$ comes from the lower limit
of the $u$-integral. To the leading order we find
\begin{equation}  \label{phi2splzanearhor}
\langle \varphi ^{2}\rangle ^{(a)}_{\beta }\approx - \frac{k_D^{D-1}z^D
\lambda _{\beta }^{\frac{D_1}{2}}e^{-2\lambda _{\beta }z_a}}{2^{D_1}\pi ^{%
\frac{D_1}{2}-1}z_a^{\frac{D_1}{2}}} \frac{I_{\nu }^2(\lambda _{\beta } z)}{%
c_a(\lambda _{\beta }z_a)} ,
\end{equation}
and the VEV is exponentially small. As it can be seen from general formula,
for the zero mode in the same limit the VEV vanishes as $z_a^{-D_1-2\nu }$.

It is also of interest to consider the behavior of the VEV for large values
of the AdS energy scale $k_D$ corresponding to strong gravitational fields.
When the values of the other parameters and the coordinates $a$ and $y$ are
fixed, for nonzero KK modes one has $\lambda _{\beta }z_a,\lambda _{\beta
}z\gg 1 $. Additionally assuming $\lambda _{\beta }|z-z_a|\gg 1$, from Eq. (%
\ref{phi2spllargelamb1}) we see that $\langle \varphi ^{2}\rangle
^{(a)}_{\beta } $ behaves as $(k_Dz)^{D-1}|z-z_a|^{-D_1/2}exp[-2\lambda
_{\beta }|z-z_a|]$ and is exponentially small. For the zero KK mode from the
general formulae we can see that the corresponding contribution to the VEV
of the field square behaves as $k_D^{D_1-1}e^{D_2k_Dy}\exp [2\nu k_D(a-y)]$
in the region $y>a$ and like $k_D^{D_1-1}e^{D_2k_Dy}\exp [(D_1+2\nu )
k_D(y-a)]$ in the region $y<a$. Note that the corresponding factors in the
VEVs integrated over the internal subspace (see Eq. (\ref{phi2integrated}))
contain an additional factor $e^{-D_2k_Dy}$ coming from the volume element
and are exponentially suppressed everywhere in the parameter space. Hence,
the strong gravitational field suppresses the vacuum fluctuations. The
similar behavior for the boundary induced quantum effects in the
gravitational field of the global monopole is described in Ref.~\cite%
{Saha03mon}.

\section{VEV in two branes geometry}

\label{sec:phi2twopl}

The branes divide the space into three distinct sections: $0<z<z_a$, $%
z_a<z<z_b$, and $z>z_b$. In general, there is no reason for the inverse
curvature radius $k_D$ to be the same in these three sections, as branes may
separate different phases of theory. The VEVs in the first and last regions
are the same as in the corresponding geometry of a single brane described in
the previous section. In this section we consider the VEV for the field
square in the region between two branes. Note that in the braneworld
scenario with two branes based on the orbifolded version of the model this
region is employed only. Taking the coincidence limit in the corresponding
formulae for the Wightman function and using the integration formula (\ref%
{rel3}), for the VEV of the field square we obtain two equivalent
representations
\begin{eqnarray}
\langle 0|\varphi ^2 |0\rangle &=&\langle \varphi ^2\rangle ^{(0)}+\langle
\varphi ^2\rangle ^{(j)} - \frac{k_{D}^{D-1}z^{D}}{2^{D_{1}-1}\pi ^{\frac{%
D_{1}}{2}}\Gamma \left( \frac{D_1}{2}\right)} \sum_{\beta }|\psi _{\beta
}(X)|^2  \notag \\
&& \times \int _{\lambda _{\beta }}^{\infty }du\, u (u^2-\lambda _{\beta
}^2)^{\frac{D_1}{2}-1}\Omega _{j\nu }(uz_{a},uz_{b}) G_{\nu
}^{(j)2}(uz_{j},uz) ,  \label{phi2twoplates}
\end{eqnarray}%
with $j=a,b$. The last term on the right of this formula is finite on the
plate at $z=z_j$ and diverges for the points on the brane $z=z_{j^{\prime}}$%
, $j^{\prime}\neq j$. These divergences are the same as those for a single
brane at $z=z_{j^{\prime}}$. As a result if we write the VEV of the field
square in the form
\begin{equation}  \label{phi2twopl1}
\langle 0|\varphi ^2 |0\rangle = \langle \varphi ^2\rangle
^{(0)}+\sum_{j=a,b}\langle \varphi ^2\rangle ^{(j)}+\langle \varphi
^2\rangle ^{(ab)},
\end{equation}
then the interference part $\langle \varphi ^2\rangle ^{(ab)}$ is finite on
both plates. In the case of one parameter manifold $\Sigma $ with the size $L
$ and for a given $k_D$, the VEV in Eq. (\ref{phi2twoplates}) is a function
on $z_b/z_a$, $L/z_a$, and $z/z_a$. Note that for the points on the brane $%
z=z_j$ in the last term on the right (\ref{phi2twoplates}) one has $G_{\nu
}^{(j)}(uz_{j},uz_{j})=-B_j$ and this term vanishes for Dirichlet boundary
conditions.

By the same way, as in the single brane case, it can be seen that in the
limit $k_D\to 0$ the result for the geometry of two parallel Robin plates in
the bulk $R^{(D_1,1)}\times \Sigma $ is obtained:
\begin{eqnarray}  \label{phi2twoplMink}
\langle \varphi ^{2}\rangle ^{(ab)}& \approx & \langle \varphi ^{2}\rangle
^{(ab)}_{R^{(D_1,1)}\times \Sigma }=\frac{(4\pi )^{- \frac{D_1}{2}}}{\Gamma
\left( \frac{D_1}{2}\right)}\sum_{\beta }\left| \psi _{\beta }(X)\right| ^2
\notag \\
&& \times \int_{\sqrt{m^2+\lambda _{\beta }^{2}}}^{\infty } du\, \frac{%
(u^2-m^2-\lambda _{\beta }^{2})^{\frac{D_1}{2}-1}}{\tilde c_{a}(u)\tilde
c_{b}(u)e^{2u(b-a)}-1}\left[ 2-\sum_{j=a,b}\frac{e^{-2u|y-j|}}{\tilde
c_{j}(u)}\right] ,
\end{eqnarray}
with the notation $\tilde c_{j}(u)$ from Eq. (\ref{cj}).

For large KK masses along $\Sigma $, $z_{a}\lambda _{\beta } \gg 1$, for the
contribution of the mode with a given $\beta $ one finds
\begin{equation}  \label{phi22pllargelamb}
\langle \varphi ^{2}\rangle ^{(ab)}_{\beta }\approx \frac{(k_Dz)^{D-1}}{%
(4\pi )^{\frac{D_1}{2}}\Gamma \left( \frac{D_1}{2}\right)} \int_{\lambda
_{\beta }}^{\infty } du\, \frac{(u^2-\lambda _{\beta }^{2})^{\frac{D_1}{2}-1}%
}{c_{a}(u z_a) c_{b}(u z_b)e^{2u(z_b-z_a)}-1}\left[ 2-\sum_{j=a,b}\frac{%
e^{-2u|z-z_j|}}{c_{j}(u z_j)}\right] ,
\end{equation}
where $\langle \varphi ^{2}\rangle ^{(ab)}_{\beta }$ is defined by the
relation similar to (\ref{phi2beta}). Additionally assuming the conditions $%
\lambda _{\beta }(z_b-z_a)\gg 1$ and $\lambda _{\beta }|z-z_j|\gg 1$, for
this quantity we obtain the formula
\begin{equation}  \label{phi22pllargelamb1}
\langle \varphi ^{2}\rangle ^{(ab)}_{\beta }\approx \frac{(k_Dz)^{D-1}}{%
(4\pi )^{\frac{D_1}{2}}} \frac{e^{-2\lambda _{\beta }(z_b-z_a)}}{%
c_{a}(\lambda _{\beta }z_a)c_{b}(\lambda _{\beta }z_b)}\frac{\lambda _{\beta
}^{\frac{D_1}{2}-1}}{(z_b-z_a)^{\frac{D_1}{2}}}.
\end{equation}
In particular, for $L_a\ll 1/k_D$ this formula takes place for all KK modes
along $\Sigma $ with nonzero masses.

Now let us consider limiting cases when the general formula for the
interference part $\langle \varphi ^2\rangle ^{(ab)}$ can be simplified.
First of all assume that $\lambda _{\beta }z_b\gg 1$ and $z_a \lambda
_{\beta }\lesssim 1$. This limit corresponds to large interbrane distances
compared with the AdS curvature radius $k_D^{-1}$ and is realized in the
braneworld scenarios for the solution of the hierarchy problem. In this
limit the main contribution into the $u$-integral comes from the region near
the lower limit and for the interference part to the leading order we have
\begin{eqnarray}
\langle \varphi ^{2}\rangle ^{(ab)}_{\beta }&\approx & -\frac{k_D^{D-1}
z^{D}\lambda _{\beta }^{\frac{D_1}{2}}e^{-2\lambda _{\beta }z_b}}{2^{D_1}
\pi ^{\frac{D_1}{2}-1}z_b^{\frac{D_1}{2}}c_{b}(\lambda _{\beta }z_b)} K_{\nu
}(\lambda _{\beta }z)  \notag \\
& & \times \frac{\bar{I}_{\nu }^{(a)}(z_a\lambda _{\beta }) }{\bar{K}_{\nu
}^{(a)}(z_a\lambda _{\beta })}\left[ \frac{\bar{I}_{\nu }^{(a)}(z_a\lambda
_{\beta })}{\bar{K}_{\nu }^{(a)}(z_a\lambda _{\beta })}K_{\nu }(\lambda
_{\beta }z)-2I_{\nu }(\lambda _{\beta }z) \right] ,  \label{phi22pllargezbn}
\end{eqnarray}
for the nonzero KK modes along $\Sigma $. In the limit $z_a, z\ll \lambda
_{\beta }^{-1}\ll z_b$ the corresponding formula takes the form
\begin{equation}  \label{phi22plas1a}
\langle \varphi ^{2}\rangle ^{(ab)}_{\beta }\approx \frac{%
k_D^{D-1}z^{D_2}(z_a\lambda _{\beta })^{2\nu }(z\lambda _{\beta })^{\frac{D_1%
}{2}}e^{-2\lambda _{\beta }z_b}}{2^{D_1+2\nu -1}\pi ^{\frac{D_1}{2}%
-1}c_{a}(\nu )c_{b}(z_b\lambda _{\beta })\Gamma ^2(\nu +1)} \left( \frac{z}{%
z_b}\right) ^{\frac{D_1}{2}} \left[1-\frac{(z_a/z)^{2\nu }}{2 c_a(\nu )}%
\right] .
\end{equation}
For $\lambda _{\beta }z\gg 1$ and $z_a \lambda _{\beta }\lesssim 1$, by
using asymptotic formulae for the Bessel modified function one finds
\begin{eqnarray}  \label{phi22plas1}
\langle \varphi ^{2}\rangle ^{(ab)}_{\beta }&\approx & \frac{%
k_D^{D-1}z^{D_2}(z\lambda _{\beta })^{\frac{D_1}{2}-1}}{2^{D_1}\pi ^{\frac{%
D_1}{2}-1}c_{b}(z_b\lambda _{\beta })} \left( \frac{z}{z_b}\right) ^{\frac{%
D_1}{2}} e^{-2\lambda _{\beta }z_b}  \notag \\
&& \times \frac{\bar{I}_{\nu }^{(a)}(z_a\lambda _{\beta })}{\bar{K}_{\nu
}^{(a)}(z_a\lambda _{\beta })}\left[ 1-\frac{e^{-2\lambda _{\beta }(z_b-z)}}{%
\lambda _{\beta }z_bc_{b}(\lambda _{\beta }z_b)}\right] .
\end{eqnarray}

And finally, in the limit $z_a \ll z_b\ll \lambda _{\beta }^{-1}$ to the
leading order we can put 0 instead of $\lambda _{\beta }$ in the lower limit
of the integral over $u$ and by using the asymptotic formulae for the Bessel
modified functions for small values of the argument, it can be seen that $%
\langle \varphi ^{2}\rangle ^{(ab)}_{\beta }\sim (z_a/z_b)^{2\nu } g(z/z_b)$%
. If in addition one has $z\ll z_b$ the following formula is obtained
\begin{eqnarray}  \label{phi22plas2}
\langle \varphi ^{2}\rangle ^{(ab)}_{\beta } &\approx & \frac{%
k_D^{D-1}z^{D_2}(z_a/z_b)^{2\nu }(z/z_b)^{D_1}}{2^{D_1+2\nu -2}\pi ^{\frac{%
D_1}{2}}\Gamma \left( \frac{D_1}{2}\right) \Gamma ^2(\nu +1)}  \notag \\
&& \times \left[ 1-\frac{(z_a/z)^{2 \nu }}{2c_a(\nu )}\right]%
\int_{0}^{\infty } du\, u^{D_1+2\nu -1} \frac{\bar{K}_{\nu }^{(b)}(u)}{\bar{I%
}_{\nu }^{(b)}(u)},
\end{eqnarray}
with the exponential suppression of the interference part. As we see for
these values of the parameters the interference part in the VEV of the field
square is mainly located near the brane at $z=z_b$. In particular, from
formula (\ref{phi22plas2}) it follows that for the zero KK mode along $%
\Sigma $ the interference part in the VEV of the field square vanishes as $%
z_b^{-D_1-2\nu }$ when the right brane tends to the AdS horizon. Note that
in the same limit the contribution of a given massive KK mode vanishes as $%
e^{-2\lambda _{\beta }z_b}/z_b^{D_1/2}$ (see formula (\ref{phi22pllargezbn}%
)). When the left brane tends to the AdS boundary, $z_a\to 0$, the
interference part vanishes as $z_a^{2\nu }$. The behavior of the
interference part in the VEV of the field square for large values of the AdS
energy scale $k_D$ directly follows from formula (\ref{phi22pllargelamb1})
in the case of nonzero KK masses and from formula (\ref{phi22plas2}) in the
case of the zero mode. Note that the corresponding contributions to the VEV
of the field square integrated over the internal space contain additional
factor $e^{-D_2k_Dy}$ coming from the volume element. In the scenario
considered in Ref. \cite{Flac03b} $L\lesssim 1/k_D$, $z_a\sim L$, and $%
z_b/z_a \gg 1$. For these values of the parameters the contribution into the
interference part for the VEV of field square to the leading order is given
by Eq. (\ref{phi22pllargezbn}) for nonzero KK modes and $\langle \varphi
^{2}\rangle ^{(ab)}_{\beta }\sim (z_a/z_b)^{2\nu } $ for the zero mode. If
in addition the observation point is far from the brane at $z=z_b$, the zero
mode contribution is determined by formula (\ref{phi22plas2}) with an
additional suppression factor $(z/z_b)^{D_1}$.

\section{An example}

\label{sec:example}

In the discussion above we have considered the general case of the internal
space. Here we take a simple example with $\Sigma =S^1$. In this case the
bulk corresponds to the $AdS_{D+1}$ spacetime with one compactified
dimension $X$. The length of this dimension we will denote by $L$. The
corresponding normalized eigenfunctions are as follows
\begin{equation}  \label{psibetS1}
\psi _{\beta }(X)=\frac{1}{\sqrt{L}}e^{2\pi i \beta X/L},\quad \lambda
_{\beta }=\frac{2\pi }{L}|\beta |,\quad \beta =0,\pm 1,\pm 2, \ldots .
\end{equation}
First of all we consider the Wightman function for the bulk without
boundaries given by formula (\ref{WAdS}). We apply to the sum over $\beta $
the Abel-Plana summation formula
\begin{equation}  \label{AbelPlana}
\sideset{}{'}{\sum}_{\beta =0}^{\infty }f(\beta )=\int _{0}^{\infty }d\beta
f(\beta ) +i\int _{0}^{\infty }d\beta \frac{f(i\beta )-f(-i\beta )}{e^{2\pi
\beta }-1},
\end{equation}
where the prime on the sum sign means that the summand $\beta =0$ should be
taken with the weight 1/2. Now it can be easily seen that the term in the
Wightman function with the first integral on the right of formula (\ref%
{AbelPlana}) corresponds to the Wightman function $\langle \varphi
(x)\varphi (x^{\prime})\rangle _{AdS_{D+1}}^{(0)}$ for the scalar field in
the bulk $AdS_{D+1}$. The latter is well investigated in literature. As a
result for the Wightman function in the bulk $AdS_{D}\times S^1$ one finds
\begin{eqnarray}  \label{WfAdSS1}
\langle \varphi (x)\varphi (x^{\prime})\rangle &=& \langle \varphi
(x)\varphi (x^{\prime})\rangle _{AdS_{D+1}}^{(0)}+ \frac{ k_{D}^{D-1}(zz^{%
\prime })^{D/2}}{2^{D-2}\pi ^{D-1}}\int d\mathbf{k}\,e^{i\mathbf{k}(\mathbf{x%
}- \mathbf{x}^{\prime })}  \notag \\
&\times & \int_{0}^{\infty }dv\,v J_{\nu }(vz)J_{\nu }(vz^{\prime })  \notag
\\
& \times & \int_{0}^{\infty }du \frac{\cosh [u(t-t^{\prime})]}{\sqrt{%
u^2+v^2+k^2}} \frac{\cosh [(X-X^{\prime})\sqrt{u^2+v^2+k^2}]}{e^{L\sqrt{%
u^2+v^2+k^2}}-1}.
\end{eqnarray}
To evaluate the corresponding VEV for the field square we take in this
formula the coincidence limit. In this limit all divergences are contained
in the part $\langle \varphi (x)\varphi (x^{\prime})\rangle
_{AdS_{D+1}}^{(0)}$, and the additional part on the right of formula (\ref%
{WfAdSS1}) induced by the compactness of a single dimension leads to the
finite result. To evaluate the corresponding integrals in the coincidence
limit, first we introduce polar coordinates $(r,\theta )$ in the plane $(k,u)
$. After the evaluation the simple $\theta $-integral, we introduce the
polar coordinates $(v,\phi )$ in the plane $(x,r)$. By using the formula
\cite{Prudnikov2}
\begin{equation}  \label{intform3}
\int_{0}^{1}dx\, x(1-x^2)^{\beta -1}J_{\nu }^2(v x)=\Gamma (\beta ) \left(
\frac{v}{2}\right) ^{2\nu }\frac{{}_{1}F_{2}\left( \nu + 1/2; \nu + 1+\beta
, 2\nu +1; -v^2\right) }{2\Gamma (\nu +1)\Gamma (\nu +1+\beta )} ,
\end{equation}
the $\phi $-integral is expressed through the hypergeometric function $%
{}_{1}F_{2}$, and one finds
\begin{eqnarray}  \label{phi2AdSS1}
\langle \varphi ^2\rangle &=& \langle \varphi ^2\rangle _{AdS_{D+1}}^{(0)} +%
\frac{2k_D^{D-1}(z/2L)^{D+2\nu }}{\pi ^{\frac{D-1}{2}}\Gamma (\nu +1)\Gamma
\left( \nu + \frac{D+1}{2}\right)}  \notag \\
& & \times \int_{0}^{\infty }du \frac{u^{D+2\nu -1}}{e^{u}-1}\,
{}_{1}F_{2}\left( \nu + \frac{1}{2}; \nu + \frac{D+1}{2}, 2\nu +1; -\frac{%
z^2u^2}{L^2}\right) .
\end{eqnarray}
As we see, unlike to the case of $AdS_{D+1}$ bulk, the VEV for the field
square in the bulk $AdS_{D}\times S^1$ is a function on $z$. Note that the
second term on the right of formula (\ref{phi2AdSS1}) is always positive.
For small values of the ratio $z/L$, to the leading order the integral is
equal to $\Gamma (D+2\nu )\zeta _{R}(D+2\nu)$, with $\zeta _{R}(x)$ being
the Riemann zeta function, and this term behaves as $(z/L)^{D+2\nu}$. For
large values of $z/L$ it can be seen that to the leading order one has
\begin{equation}  \label{phi2S1largez}
\langle \varphi ^2\rangle \approx \frac{\zeta _{R}(D-1)}{2\pi ^{\frac{D+1}{2}%
}} \Gamma \left( \frac{D-1}{2}\right) \left( \frac{k_Dz}{L}\right) ^{D-1} ,
\end{equation}
where we have taken into account that the quantity $\langle \varphi
^2\rangle _{AdS_{D+1}}^{(0)}$ does not depend on $z$. Hence, near the AdS
horizon the VEV of the field square is dominated by the second term on the
right of Eq. (\ref{phi2AdSS1}). Note that in the limit $k_D\to 0$ the VEV $%
\langle \varphi ^2\rangle _{AdS_{D+1}}^{(0)}$ vanishes, and taking into
account that $zk_D\to 1$, from (\ref{phi2S1largez}) we obtain the standard
result for the VEV in the bulk $R^{(D-1,1)}\times S^1$.

\begin{figure}[tbph]
\begin{center}
\begin{tabular}{cc}
\epsfig{figure=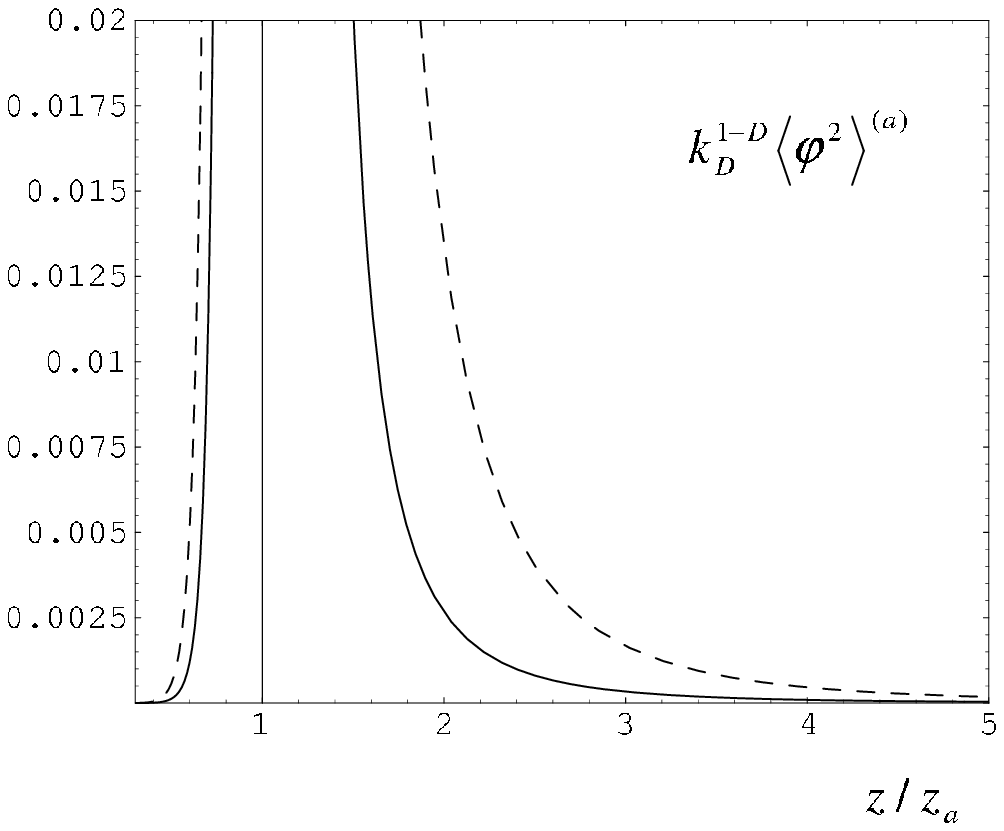,width=6.5cm,height=6cm} & \quad %
\epsfig{figure=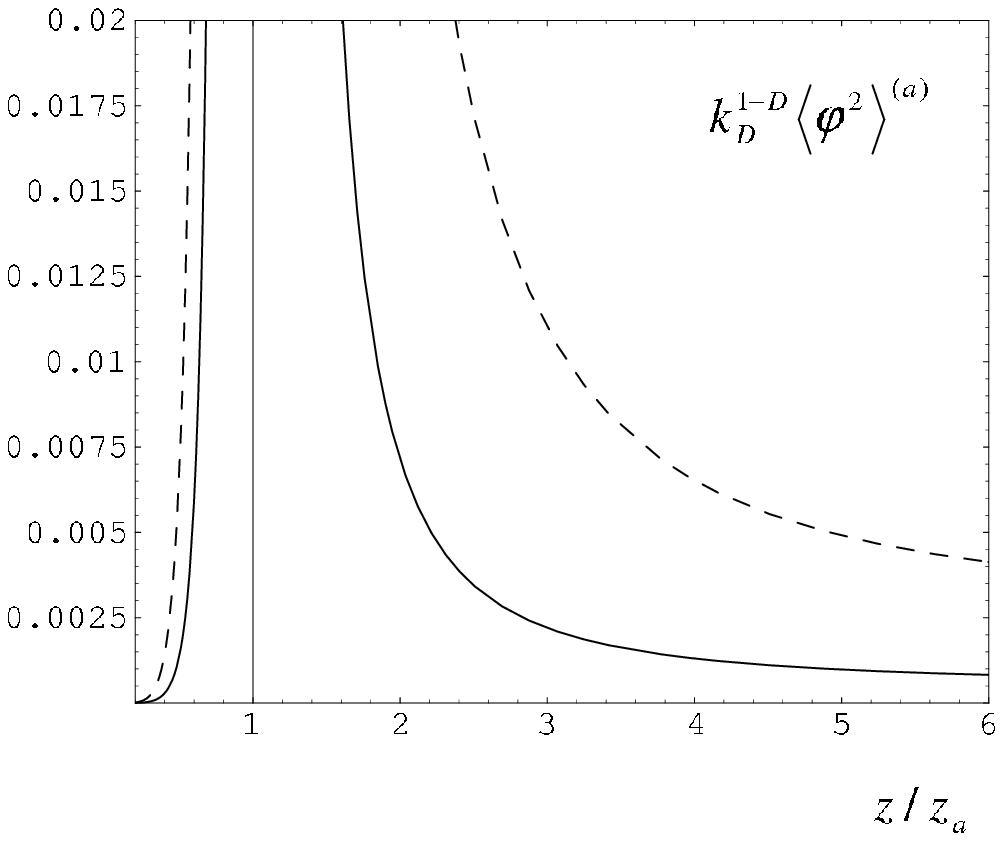,width=6.5cm,height=6cm}%
\end{tabular}%
\end{center}
\caption{Single brane induced VEV of the field square as a function on $z/z_a
$ for $D=5$ minimally (left panel) and conformally (right panel) coupled
massless scalars with $A_a/B_a=-0.4$. The full and dashed curves correspond
to the values $L/z_a=1$ and $L/z_a=0.2$ respectively.}
\label{fig1}
\end{figure}

Brane induced VEVs of the field square for the geometry under consideration
are obtained from general formulae (\ref{phi2spl}), (\ref{phi2twoplates}) by
the replacements
\begin{equation}  \label{replS1}
\sum_{\beta }|\psi _{\beta }(X)|^2\to \frac{2}{L}\sideset{}{'}{\sum}_{\beta
=0}^{\infty }, \quad \lambda _{\beta }\to \frac{2\pi }{L}|\beta |, \quad
D_1\to D-1 .
\end{equation}
As it has been mentioned in section \ref{sec:vevphi2}, for $D_2=1$ and
conformally coupled massless scalar field ($\nu =1/2$) the single brane
induced VEV is finite on the AdS horizon. The corresponding limiting value
is obtained from (\ref{phi2splsmallza2}):
\begin{equation}  \label{phi2S1onhor}
\langle \varphi ^2\rangle ^{(a)}|_{\nu =1/2,z\to \infty } =-\frac{%
k_D^{D-1}\Gamma (D/2)z_a}{2^{D-2}\pi ^{\frac{D}{2}}c_a(1/2)L}.
\end{equation}
For the general case, the VEV in this limit behaves as $z^{1-2\nu }$. In
figure \ref{fig1} we have plotted the dependence of the single brane induced
VEV $\langle \varphi ^2\rangle ^{(a)}$ on the ratio $z/z_a$ for $D=5$
minimally (left panel) and conformally (right panel) coupled massless scalar
fields with $A_a/B_a=-0.4$. The full and dashed curves correspond to the
values $L/z_a=1$ and $L/z_a=0.2$ respectively. As it has been shown in
section \ref{sec:vevphi2}, in the limit $z\to 0$ the VEV behaves as $%
z^{D+2\nu }$. For the points near the AdS horizon, $z\to \infty $, the main
contribution comes from the zero mode and the VEV behaves as $z^{1-2\nu }$.
On the brane surface the VEV diverges like $|z-z_a|^{2-D}$.

\begin{figure}[tbph]
\begin{center}
\begin{tabular}{cc}
\epsfig{figure=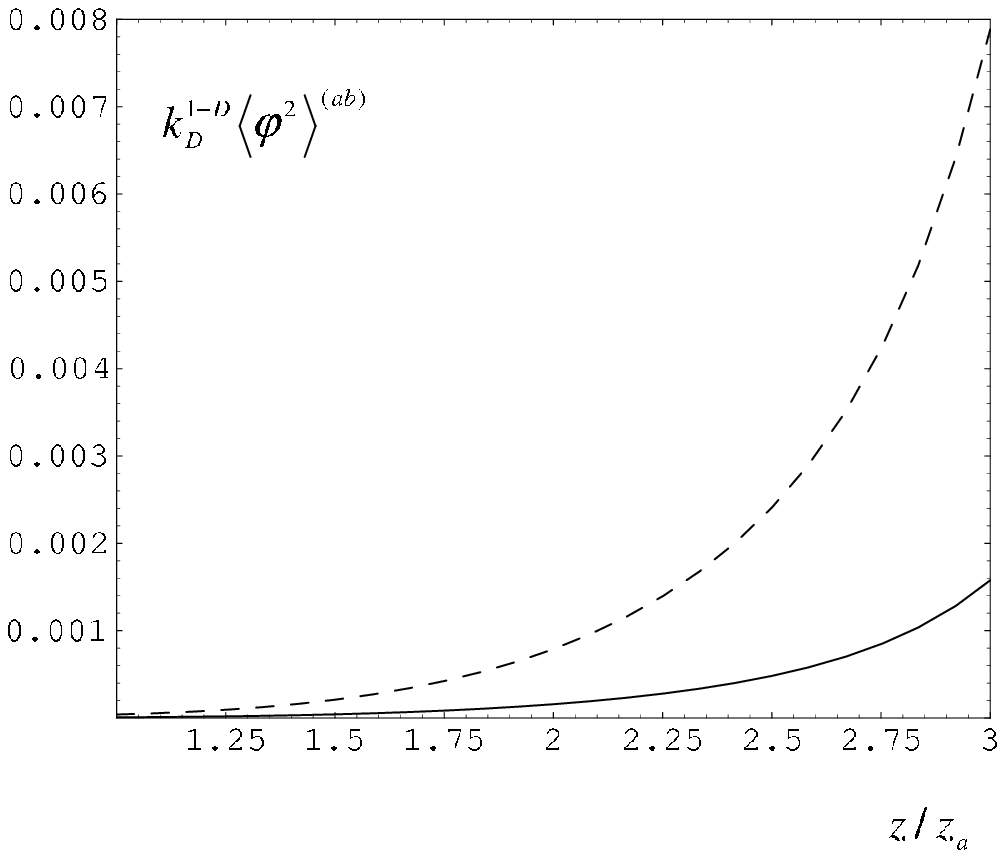,width=6.5cm,height=6cm} & \quad %
\epsfig{figure=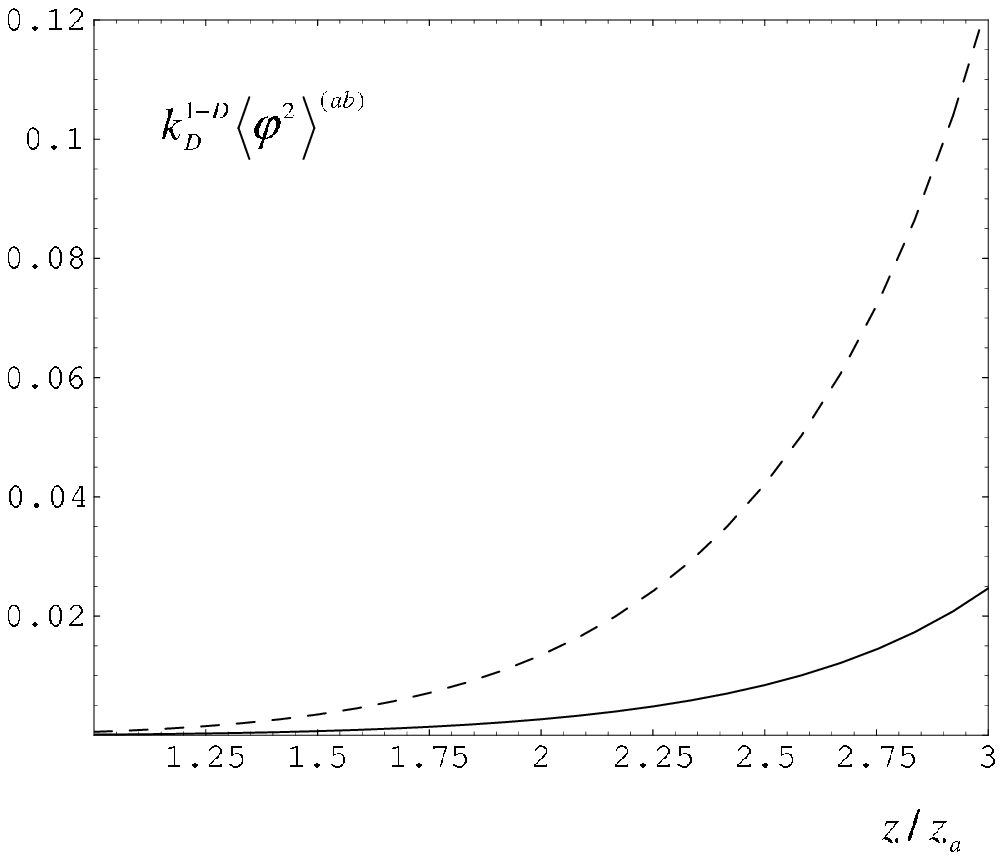,width=6.5cm,height=6cm}%
\end{tabular}%
\end{center}
\caption{Interference part in the VEV of the field square as a function on $%
z/z_a$ for $D=5$ minimally (left panel) and conformally (right panel)
coupled massless scalars with $A_j/B_j=-0.4$, $j=a,b$. The full and dashed
curves correspond to the values $L/z_a=1$ and $L/z_a=0.2$ respectively. The
interbrane distance corresponds to the value $z_b/z_a=3$.}
\label{fig2}
\end{figure}

In figure \ref{fig2} we present the graphs for the interference part in the
VEV of the field square as a function on $z/z_a$ for $D=5$ minimally (left
panel) and conformally (right panel) coupled massless scalars with $%
A_j/B_j=-0.4$, $j=a,b$. The graphs are plotted for the interbrane
distance corresponding to $z_b/z_a=3$. Note that in accordance
with the general discussion in the previous section one has
$\langle \varphi ^2\rangle ^{(ab)}_{z=z_a}\ll \langle \varphi
^2\rangle ^{(ab)}_{z=z_b}$.

It is not difficult to generalize the corresponding results for the case of
the internal space $\Sigma =S^{D_2}$ with radius $R_0$ and curvature scalar $%
R_{(\gamma )}=-D_2(D_2-1)/R_0^2$. Now the eigenfunctions $\psi _{\beta }(X)$
are expressed in terms of spherical harmonics of degree $l$, $l=0,1,2,\ldots
$. The VEVs for the field square are obtained from the general formulae in
sections \ref{sec:vevphi2} and \ref{sec:phi2twopl} by the replacements
\begin{eqnarray}
\sum_{\beta }|\psi _{\beta }(X)|^2 &\to & \frac{\Gamma \left( \frac{D_2+1}{2}%
\right)}{2 \pi ^{\frac{D_2+1}{2}}R_0^{D_2}}\sum_{l=0}^{\infty } (2l+D_2-1)%
\frac{\Gamma (l+D_2- 1)}{l! \Gamma (D_2)},  \label{SD1psi} \\
\lambda _{\beta } &\to & \frac{1}{R_0}\sqrt{l(l+D_2-1)+\zeta D_2(D_2-1)} ,
\label{SD1lambda}
\end{eqnarray}
where the factor under the sum sign on the right in Eq. (\ref{SD1psi}) is
the degeneracy of each angular mode with a given $l$.

\section{Conclusion}

From the point of view of embedding the braneworld model into a more
fundamental theory, such as string/M-theory, one may expect that a more
complete version of this scenario must admit the presence of additional
extra dimensions compactified on a manifold $\Sigma $. In the present paper
we have extended the previous work describing the local vacuum effects in
the braneworlds with the AdS bulk on a higher dimensional brane models which
combine both the compact and warped geometries. This problem is also of
separate interest as an example with gravitational, topological, and
boundary polarizations of the vacuum, where one-loop calculations can be
performed in closed form. We have investigated the Wightman function and the
vacuum expectation value of the field square for a scalar field with an
arbitrary curvature coupling parameter satisfying Robin boundary conditions
on two parallel branes in $AdS_{D_1+1}\times \Sigma $ spacetime. The KK
modes corresponding to the radial direction are zeros of a combination of
the cylinder functions. The application of the generalized Abel-Plana
formula to the corresponding mode sum in the expression of the Wightman
function allowed us to extract the boundary-free part and to present the
brane induced parts in terms of integrals exponentially convergent in the
coincidence limit of the arguments. In the region between two branes the
Wightman function is presented in two equivalent forms given by Eqs. (\ref%
{W15}) and (\ref{W17}). The first term on the right of these formulae is the
Wightman function for the bulk without boundaries. The second one is induced
by a single brane and the third term is due to the presence of the second
brane. Further we give an application of our results for two branes case to
the higher dimensional version of the Randall-Sundrum braneworld with
arbitrary mass terms on the branes. For the untwisted scalar the Robin
coefficients are expressed through these mass terms and the curvature
coupling parameter by formulae (\ref{AtildeRS}). For the twisted scalar
Dirichlet boundary conditions are obtained on both branes.

The expectation values for the field square are obtained from the Wightman
function taking the coincidence limit. For the case of a single brane
geometry this leads to formula (\ref{phi2spl}) for the region $z>z_a$. The
corresponding formula in the region $z<z_a$ is obtained from Eq. (\ref%
{phi2spl}) by replacements $I_{\nu }\rightleftarrows K_{\nu }$. For a one
parameter manifold $\Sigma $, the VEV of the field square induced by a
single brane is a function on the proper distance of the observation point
from the brane and on the ratio of the physical size of $\Sigma$ (from the
viewpoint of an observer residing on the brane) to the AdS curvature radius.
As a partial check of our formulae, we have shown that in the limit when the
AdS radius goes to infinity the result for the brane in the bulk $%
R^{(D_1,1)}\times \Sigma $ is obtained. On the boundary the VEV for the
field square diverges. In the limit when the proper distance from the brane
is much smaller compared with the AdS curvature radius and proper size $L_a$
of the internal manifold, the leading term of the corresponding asymptotic
expansion for a contribution of a given KK mode along $\Sigma $ is given by
expression (\ref{phi2splnear}). This term does not depend on the Robin
coefficient and has different signs for Dirichlet and non-Dirichlet scalars.
The coefficients for the subleading asymptotic terms will depend on the bulk
curvature radius, Robin coefficient and on the mass of the field. In the
limit when the KK masses $\lambda _{\beta }^{(a)}$ are much larger compared
to the AdS energy scale $k_D$, the contribution of a given mode with fixed $%
\beta $ to the VEV of the field square is given by formula (\ref%
{phi2spllargelamb1}). From this formula it follows that if additionally one
has $\lambda _{\beta }^{(a)}|y-a|\gg 1$, the contribution of the
corresponding KK modes is exponentially suppressed. In the region $y>a$, for
large distances from the brane, $k_D(y-a)\gg 1$, and small KK masses $%
\lambda _{\beta }^{(a)}$ compared with AdS energy scale, $\lambda _{\beta
}^{(a)}\ll k_D$, the contribution of a given KK mode along the internal
space $\Sigma $ is approximated by formula (\ref{phi2splsmallza}). In both
subcases $k_De^{-k_D(y-a)}\ll \lambda _{\beta }^{(a)}$ and $%
k_De^{-k_D(y-a)}\gg \lambda _{\beta }^{(a)}$ we have exponentially
suppressed VEVs by the factors $\exp [-2e^{k_D(y-a)}\lambda _{\beta
}^{(a)}/k_D]$ and $\exp [2\nu (a-y)]$ in the first and second cases
respectively. For large distances from the brane in the region $y<a$, the
contribution of a given KK mode is presented in the form (\ref{phi2splsmallz}%
). This formula is further simplified in the subcases of large and small KK
masses $\lambda _{\beta }^{(a)}$. In the case of large mass, $%
e^{k_D(y-a)}\ll k_D/\lambda _{\beta }^{(a)}\ll 1$, one has an exponential
suppression by the factor $\exp [-2\lambda _{\beta }^{(a)}/k_D]$. In the
limit of small masses, $\lambda _{\beta }^{(a)}\ll k_D$, the suppression is
by the factor $\exp [k_D(y-a)(D_1+2\nu )]$. On the AdS horizon ($z\to \infty
$), the brane induced VEV is exponentially small for nonzero KK masses along
$\Sigma $ and behaves as $z^{D_2-2\nu }$ for the zero mode. On the AdS
boundary, corresponding to $z=0$, the brane induced VEV vanishes as $%
z^{D+2\nu }$. When the brane position tends to the AdS boundary, $z_a\to 0$,
the brane induced VEV in the region $z>z_a$ vanishes as $z_a^{2\nu }$. In
the limit when the brane tends to the AdS horizon, $z_a\to \infty $, the VEV
in the region $z<z_a$ behaves as $\exp[-2\lambda _{\beta }z_a]/z_a^{D_1/2}$
for massive KK modes along $\Sigma $, and vanishes as $z_a^{-D_1-2\nu }$ for
the zero mass mode. For large values of the AdS energy scale $k_D$
corresponding to strong gravitational fields, the VEVs integrated over the
internal space (see formula (\ref{phi2integrated})) are exponentially
suppressed everywhere in the parameter space.

For two branes geometry, by using the corresponding expression for the
Wightman function, VEV for the field square in the region between the branes
is presented as a sum of boundary-free, single brane induced and
interference parts, Eq. (\ref{phi2twopl1}). The latter is regular everywhere
including the points on the branes. In the limit $k_D\to 0$ the result for
the geometry of two branes in the bulk $R^{(D_1,1)}\times \Sigma $ is
obtained. For large KK masses along $\Sigma $, $\lambda _{\beta }^{(a)}\gg
k_D$, the contribution of the KK mode with a given $\beta$ is determined by
formula (\ref{phi22pllargelamb}). This formula is further simplified when
the branes are not too close to each other and the observation point is
sufficiently far from the branes, $\lambda _{\beta }(z_b-z_a)\gg 1$ and $%
\lambda _{\beta }(z-z_j)\gg 1$, $j=a,b$. In this case we obtain formula (\ref%
{phi22pllargelamb1}) with exponentially small interference part. For the
modes with masses in the range $z\ll \lambda _{\beta }^{-1}\ll z_b$, the
contribution to the VEV of the field square is determined by formula (\ref%
{phi22plas1a}) with two small factors, $e^{-2\lambda _{\beta }z_b}$ and $%
(z/z_b)^{D_1/2}$. For small values of KK masses, $z_a\ll z_b\ll \lambda
_{\beta }^{-1}$, and large interbrane distances, to the leading order the
contribution of a given KK mode does not depend on the KK mass and the
corresponding VEV is exponentially suppressed by the factor $\exp [2\nu k_D
(a-b)]$. If in addition the observation point is far from the brane at $y=b$%
, additional suppression factor $\exp [D_1k_D(y-b)]$ appears and the
interference part is mainly located near the brane $y=b$. When the right
brane position tends to the AdS horizon, $z_b\to \infty $, the interference
part in the VEV of the field square vanishes as $z_b^{-D_1-2\nu }$ for the
zero KK mode along $\Sigma $. For a given massive KK mode this quantity
behaves as $\exp [-2\lambda _{\beta }z_b]/z_b^{D_1/2}$. When the left brane
tends to the AdS boundary, the interference part vanishes as $z_a^{2\nu }$.

As an illustration of general results, in section \ref{sec:example} we
consider an example of the internal space $\Sigma =S^1$. Using the
Abel-Plana formula, the Wightman function for the boundary-free geometry is
presented in the form (\ref{WfAdSS1}), where the first term on the right is
the corresponding function for the $AdS_{D+1}$ bulk. The VEV of the field
square is obtained in the coincidence limit and has the form (\ref{phi2AdSS1}%
). Unlike to the case of $AdS_{D+1}$ bulk, this VEV for the bulk $%
AdS_{D}\times S^1$ depends on $z$ coordinate and behaves as $(z/L)^{D+2\nu }$
for small values of the ratio $z/L$ and as $(z/L)^{D-1}$ for large values of
this ratio. In figures \ref{fig1} and \ref{fig2} we have presented single
brane induced and interference parts in the VEV of the field square for
minimally and conformally coupled scalars in the bulk geometry $%
AdS_{5}\times S^1$ as functions on the distance from the brane at $y=a$. We
also discuss the generalization of the corresponding results for the
internal space $S^{D_2}$. The VEVs in this case are obtained from the
general formulae in sections \ref{sec:vevphi2} and \ref{sec:phi2twopl} by
the replacements (\ref{SD1psi}) and (\ref{SD1lambda}). In particular, from
the point of view of embedding the Randall-Sundrum model into the string
theory, the case of the bulk $AdS_{5}\times S^5$ is of special interest.
This spacetime also plays an important role in the discussions of the
holographic principle.

Note that in this paper we have considered boundary induced vacuum
densities which are finite away from the boundaries. We expect
that similar results would be obtained in the model where instead
of externally imposed boundary condition the fluctuating field is
coupled to a smooth background potential that implements the
boundary condition in a certain limit \cite{Grah02}.

\section*{Acknowledgments}

The work was supported by the CNR-NATO Senior Fellowship, by ANSEF Grant No.
05-PS-hepth-89-70, and in part by the Armenian Ministry of Education and
Science, Grant No. 0124. The author acknowledges the hospitality of the
Abdus Salam International Centre for Theoretical Physics (Trieste, Italy)
and Professor Seif Randjbar-Daemi for his kind support.

\end{document}